\documentclass[twocolumn]{aastex6}

\usepackage{float}
\usepackage{mathtools}
\usepackage{enumitem}
\usepackage{bm}
\usepackage{amsmath}
\usepackage{amssymb}
\usepackage{graphicx}
\usepackage{hyperref}
\usepackage{natbib}

\shorttitle{LSS Projection Effect}
\shortauthors{Yuan et. al.}
\begin{document}
\hypersetup{
 colorlinks=true,
 linkcolor=blue,
 anchorcolor=red,
 citecolor=cyan,
 urlcolor=magenta}

\newcommand\myemail[1]{\href{mailto:#1}{{\nolinkurl{#1}}}}
\defcitealias{Fan2010}{F10}

\title{Projection effects of large-scale structures on weak-lensing peak
abundances}

\author{Shuo Yuan\altaffilmark{1},
Xiangkun Liu\altaffilmark{2}, 
Chuzhong Pan\altaffilmark{3,1}, 
Qiao Wang\altaffilmark{4}, 
Zuhui Fan\altaffilmark{1}}

\altaffiltext{1}{Department of Astronomy, School of Physics, Peking University, Beijing 100871, China;\\ \myemail{yuanshuoastro@gmail.com}, \ \myemail{fanzuhui@pku.edu.cn}}
\altaffiltext{2}{South-Western Institute for Astronomy Research, Yunnan University, Kunming 650500, China}
\altaffiltext{3}{Center for High Energy Physics, Peking University, Beijing 100871, China}
\altaffiltext{4}{Key Laboratory for Computational Astrophysics, The Partner Group of Max Planck Institute for Astrophysics,
National Astronomical Observatories, Chinese Academy of Sciences, Beijing 100012, China;}

\begin{abstract}

High peaks in weak lensing (WL) maps originate dominantly from the
lensing effects of single massive halos. Their abundance is therefore
closely related to the halo mass function and thus a powerful cosmological
probe. On the other hand, however, besides individual massive halos,
large-scale structures (LSS) along lines of sight also contribute
to the peak signals. In this paper, with ray tracing simulations,
we investigate the LSS projection effects. We show that for current
surveys with a large shape noise, the stochastic LSS effects are subdominant.
For future WL surveys with source galaxies having a median redshift
$z_{\mathrm{med}}\sim1$ or higher, however, they are significant. For the cosmological
constraints derived from observed WL high peak counts, severe biases
can occur if the LSS effects are not taken into account properly.
We extend the model of \citet{Fan2010} by incorporating the  LSS
projection effects into the theoretical considerations. By comparing with simulation
results, we demonstrate the good performance of the improved model
and its applicability in cosmological studies. 

\end{abstract}

\section{\label{sec:intro}Introduction }

Being an important cosmological probe, the weak lensing (WL) effect
is one of the key science drivers for a number of ongoing and future
large surveys \citep[e.g.,][]{Albrecht2006,Weinberg2013,FuFan2014,Kilbinger2015,LSST2012,Amendola2013,DES2016}.
Unlike strong lensing effects where individual lens systems can be
investigated, WL analyses are statistical in nature. Therefore it
is important to explore different statistics to enrich the cosmological
gains from WL data.

The cosmic shear two-point (2pt) correlation/power spectrum analyses
are the most widely studied ones in WL cosmology \citep[e.g.,][]{Kilbinger2013,Becker2016,Hildebrandt2016}.
On the other hand, WL signals have reached the non-linear scales and
thus the 2pt statistics cannot uncover the full cosmological information
therein. The three-point correlation measurements are done for a number
of surveys with the realization of much more complications, both observational
and theoretical, than that of 2pt correlations \citep{Pen2003,Semboloni2011,Fu2014}.
Concentrating on high signal regions, WL peak statistics has emerged
as another promising means to probe non-linear structures and cosmology,
complementary to cosmic shear correlation analyses 
(\citealt{Shan2012,Shan2014}; {\color{cyan}{X. K.} }\citealt{Liu2015}; {\color{cyan}{J.}} \citealt{LiuJia2015}; \citealt{DESpeak2016}; 
 \citealt{2018MNRAS.474..712M}; \citealt{2018MNRAS.474.1116S}).

WL peaks, particularly high peaks, arise primarily from the lensing
effects of massive halos along their lines of sight \citep{White2002,Hamana2004,DietHart2010,Fan2010,Yang2011,Lin2015,Shirasaki2015}.
The high peak abundance is thus a reflection of the halo mass function,
and a sensitive cosmological probe considering further the cosmology-dependent
lensing kernel. It is less affected by baryonic physics than normal
cluster abundances where certain baryonic observable-mass relations
are needed. However, apart from massive halos, other effects can also
impact the peak signals, notably the projection effect of large-scale
structures (LSS) and the shape noise resulting from the intrinsic ellipticities
of source galaxies \citep{2005ApJ...635...60T,Fan2010,Yang2011,Hamana2012,LiuJia2016}.
To predict accurately the WL peak abundance for cosmological studies,
we should take them into account carefully.

In principle, numerical simulations can include various effects, and
we can build empirical templates of peak counts for different cosmological
models incorporating different observational effects with respect
to specific surveys \citep{DietHart2010}. By comparing with observational
peak counts, we therefore are able to derive cosmological constraints 
({\color{cyan}J.} \citealp{LiuJia2015}; \citealp{DESpeak2016}).

Such an approach is numerically intensive noting the high
dimensions of the cosmological parameter space and different astrophysical
and observational effects. Thus theoretical models are highly desirable
for performing cosmological studies efficiently. Furthermore, the
physical picture related to WL peaks can be seen more clearly in theoretical
models which need to specify different effects explicitly \citep{Marian2009,Maturi2010,Hamana2012,Lin2015,Shirasaki2015}.

In \citet{Fan2010} (here after, \citetalias{Fan2010})
the WL high peak abundance is
modeled by assuming that the true WL peaks are from the lensing effects
of individual massive halos. In addition, the shape noise effect is
carefully included, which not only generates false peaks but also
influences the peak signals from halos. The comparison with simulations
shows that the model works very well in the case of surveys with the
source galaxies having a shallow redshift $z\sim0.7$, a number density
$n_{g}\sim10\,\hbox{arcmin}^{-2}$ and a survey area $\sim150\,\hbox{deg}^{2}$.
For such surveys, the projection effect of LSS
 is minor comparing to the shape noise. This model has been applied
in the cosmological studies by analyzing WL peak counts using data
from KiDS survey (\citealt{2018MNRAS.474.1116S}  ),
CFHTLenS survey ({\color{cyan}X. K. }\citealp{Liu2016}) and CFHT Stripe 82 survey 
(\citealp{Shan2014}; {\color{cyan}X. K. }\citealp{Liu2015}).

For the ongoing and upcoming surveys, the survey depth can be improved considerably
to detect more faraway galaxies for WL analyses. This will result
in a suppression of the shape noise as well as a growth of the
LSS projection effects. In such cases, the LSS effects must be included
in the theoretical modelling carefully. In addition, the sky coverage
will be enlarged by orders of magnitude and the statistical errors of
WL peak counts will expectedly decrease. Therefore even in the case
that the LSS projection effect is minor, it is still necessary to consider this effect for
accurate modelling.

Recently, the comparisons between WL peak counts from a large set
of simulations and from the halo-based Monte Carlo model named
\texttt{CAMELUS} \citep{Lin2015} are shown in \citet{ZM2016}. It
is found that for high peaks, \texttt{CAMELUS} works well for the
source galaxies at $z_{\mathrm{s}}=1$ and with the cosmological parameters
$\Omega_{m}$ and $\sigma_{8}$ close to the current best values, where $\Omega_m$ and $\sigma_8$ are the dimensionless matter density of the universe at present and the linear extrapolated density perturbations smoothed over a top-hat scale of $8.0\,h^{-1}\mathrm{Mpc}$, respectively.
For higher values e.g., $\Omega_{m}\sim0.5$ and $\sigma_{8}\sim0.9$,
the deviations between the results from simulations and those from
\texttt{CAMELUS} are shown up. We note that for high peaks, \texttt{CAMELUS}
is essentially the same as that of \citetalias{Fan2010} and it does not include the LSS
contributions beyond halos. For high $\Omega_{m}$ and $\sigma_{8}$,
we expect stronger LSS projection effects than those of low $\Omega_{m}$
and $\sigma_{8}$. This should at least partly explain the differences
of the WL high peak counts seen in \citet{ZM2016}.

In this paper, we investigate in detail the LSS projection effect
on WL high peak counts, and improve the model of \citetalias{Fan2010} by taking the
projection effect into the theoretical considerations. We perform
extensive tests using numerical simulations, demonstrating the applicability
of the improved model for future WL studies.

The rest of the paper is organized as follows. 
\S2 presents the WL peak analyses, and the improved model for high peak abundances including the LSS projection effects. 
In \S3, we show the simulation tests in detail and validate the model performance for different survey settings. 
Summary and discussions are given in \S4.

\section{\label{sec:modelling}Modelling weak lensing peak abundance including the LSS projection effect}

\subsection{Weak gravitational lensing effect}

Photons are subjected to the gravity of cosmic structures, and deflected
when they propagate toward us. As a result, the observed images differ
from their original ones. This phenomenon is referred to as the gravitational
lensing effect. In the WL regime, the effect leads to small changes
of size and shape of the images.

Theoretically, the WL effect can be described by the lensing potential
$\phi$. Its gradient gives rise to the deflection angle, and the
second derivatives are related directly to the observational consequence
of the lensing effect. Specifically, the convergence $\kappa$ and
the shear $\gamma$, characterizing the size and the shape changes,
respectively, are given by \citep{BartSch2001}

\begin{equation}
\kappa=\frac{1}{2}\nabla^{2}\phi,\label{kappa}
\end{equation}

\begin{equation}
\gamma_{1} = \frac{1}{2} \bigg(\frac{\partial^{2}\phi}{\partial^{2}x_{1}}-\frac{\partial^{2}\phi}{\partial^{2}x_{2}}\bigg),\quad\gamma_{2}=\frac{\partial^{2}\phi}{\partial x_{1}\partial x_{2}},\label{gamma}
\end{equation}
where $\boldsymbol{x}=(x_{1},x_{2})$ is the two-dimensional angular
vector. Under the Born approximation, the convergence $\kappa$ is
the projected density fluctuation weighted by the lensing kernel,
and given by

\begin{multline}
\kappa(\boldsymbol{x})=  \frac{3H_{0}^{2}\Omega_{m}}{2} \int_{0}^{\chi_{H}}\mathrm{d}\chi' \int_{\chi'}^{\chi_{H}} \mathrm{d}\chi  \\ \bigg[p_{\mathrm{s}}(\chi)\frac{f_{K}(\chi-\chi')f_{K}(\chi')}{f_{K}(\chi)a(\chi')}\bigg]{\delta[f_{K}(\chi')\boldsymbol{x},\chi']},\label{born}
\end{multline}
where $\chi$ is the comoving radial distance, $\chi_{H}=\chi(z=\infty)$,
$a$ is the cosmic scale factor, $f_{K}$ is the comoving angular
diameter distance, $\delta$ is the 3-D density fluctuation, and $p_{\mathrm{s}}$
is the source distribution function. The cosmological parameters $H_{0}$
is the Hubble constant.

Define the lensing window function as:
\begin{equation}
w(\chi')=\int_{\chi'}^{\chi_{H}}\mathrm{d}\chi p_{\mathrm{s}}(\chi)\frac{f_{K}(\chi-\chi')}{f_{K}(\chi)},
\end{equation}
the corresponding power spectrum of $\kappa$ is then

\begin{equation}
C_{\ell}=\frac{9H_{0}^{4}\Omega_{m}^{2}}{4}\int_{0}^{\chi_{H}}\mathrm{d}\chi'\frac{w^{2}(\chi')}{a^{2}(\chi')}P_{\delta} \left( \frac{\ell}{f_{K}(\chi')},\chi'\right),\label{eq:Limb}
\end{equation}
where $P_{\delta}$ is the power spectrum of 3-D matter density perturbations.

Observationally, the brightness quadrupole moment tensor of a source 
galaxy can be measured, and from that, the source ellipticity can be extracted.
The WL effect on observed images can then be described by the Jacobian
matrix of the lensing equation, which reads

\begin{equation}
\begin{split}
\mathbf{A} & = 
 \left(
\begin{array}{cc}
1-\kappa -\gamma_1 & -\gamma_2 \\
-\gamma_2 & 1-\kappa +\gamma_1
\end{array}     \right) \\ & =  
(1-\kappa)
\left(
\begin{array}{cc}
1-g_1 & -g_2 \\
-g_2 & 1+g_1
\end{array} \right),\label{jacobian}
\end{split}
\end{equation}
where $g_{i}=\gamma_{i}/(1-\kappa)$ is the reduced shear. Considering
the intrinsic ellipticity of a source galaxy, the observed ellipticity
written in the complex form \citep{Seitz1997} is

\begin{equation}
\boldsymbol{\epsilon}= 
\begin{cases}
\dfrac{\boldsymbol{\epsilon}_\mathrm{s}+\boldsymbol{g}}{1+\boldsymbol{g^{*}}\boldsymbol{\epsilon}_\mathrm{s}}; & \text{for \ensuremath{\vert{\boldsymbol{g}}\vert \leqslant 1}  } \\
\\
\dfrac{1+\boldsymbol{g}\boldsymbol{\epsilon}_\mathrm{s}^{*}}{\boldsymbol{\epsilon}_\mathrm{s}^{*}+\boldsymbol{g^{*}}}, & \text{for \ensuremath{\vert{\boldsymbol{g}}\vert>1} }
\end{cases} \label{eobs}
\end{equation}
where $\boldsymbol{\epsilon}$ and $\boldsymbol{\epsilon}_\mathrm{s}$ are
the observed and the intrinsic ellipticities of a source, respectively.
The symbol $^{*}$ represents the complex conjugate operation. It
is seen that the observed ellipticity is closely related to the WL
shear. For $\kappa\ll1$, $\boldsymbol{\epsilon}\approx\boldsymbol{\epsilon}_\mathrm{s}+\boldsymbol{\gamma}$.
Without considering the intrinsic alignments, the correlation analyses
of $\boldsymbol{\epsilon}$ can thus give rise directly to an estimate
of the WL shear correlation \citep[e.g.,][]{Fu2008,Kilbinger2013,Fu2014,Jee2016,Becker2016,Hildebrandt2016}.

Alternatively, because of the physical relation between the shear
and the convergence as seen in Eq.(\ref{kappa}) and Eq. (\ref{gamma}),
it is possible to reconstruct the convergence $\kappa$ field from
the observed ellipcities after a suitable smoothing \citep[e.g.,][]
{Kaiser1993,Seitz1995,Bartelmann1995,Squires1996,Jullo2014}.
We note that $\kappa$ is the weighted projection of the density fluctuations
and thus the structures can be better seen visually in the $\kappa$
field than that in the shear field. Comparing to previous observations
targeting at individual clusters, the current survey cameras have
a large field of view, typically $\sim 1^{\circ} \times 1 ^{\circ}$, and thus
the boundary effects on the convergence reconstruction can be in good
control. Cosmological studies using the reconstructed convergence
fields have been carried out for different WL surveys 
(e.g., \citealp{Shan2012,Waerbeke2013,Shan2014}; {\color{cyan}J. }\citealp{LiuJia2015}; {\color{cyan}X. K. }\citealp{Liu2015,Liu2016}).

In this paper, we concentrate on WL peaks identified in convergence
fields, and particularly study the LSS projection effects on high
peak abundances.

\subsection{WL high peak abundance with stochastic LSS}

Physically, the WL convergence field reflects the projected density
distribution weighted by the lensing kernel. Peaks there should correspond
to the projected mass concentrations. Studies show that for a high
peak, its signal is primarily contributed by a single massive halo
located in the line of sight (e.g., \citealt{Yang2011}; {\color{cyan}X. K. }\citealp{Liu2014,LiuJia2016}).

\begin{figure*}
\includegraphics[width=9cm]{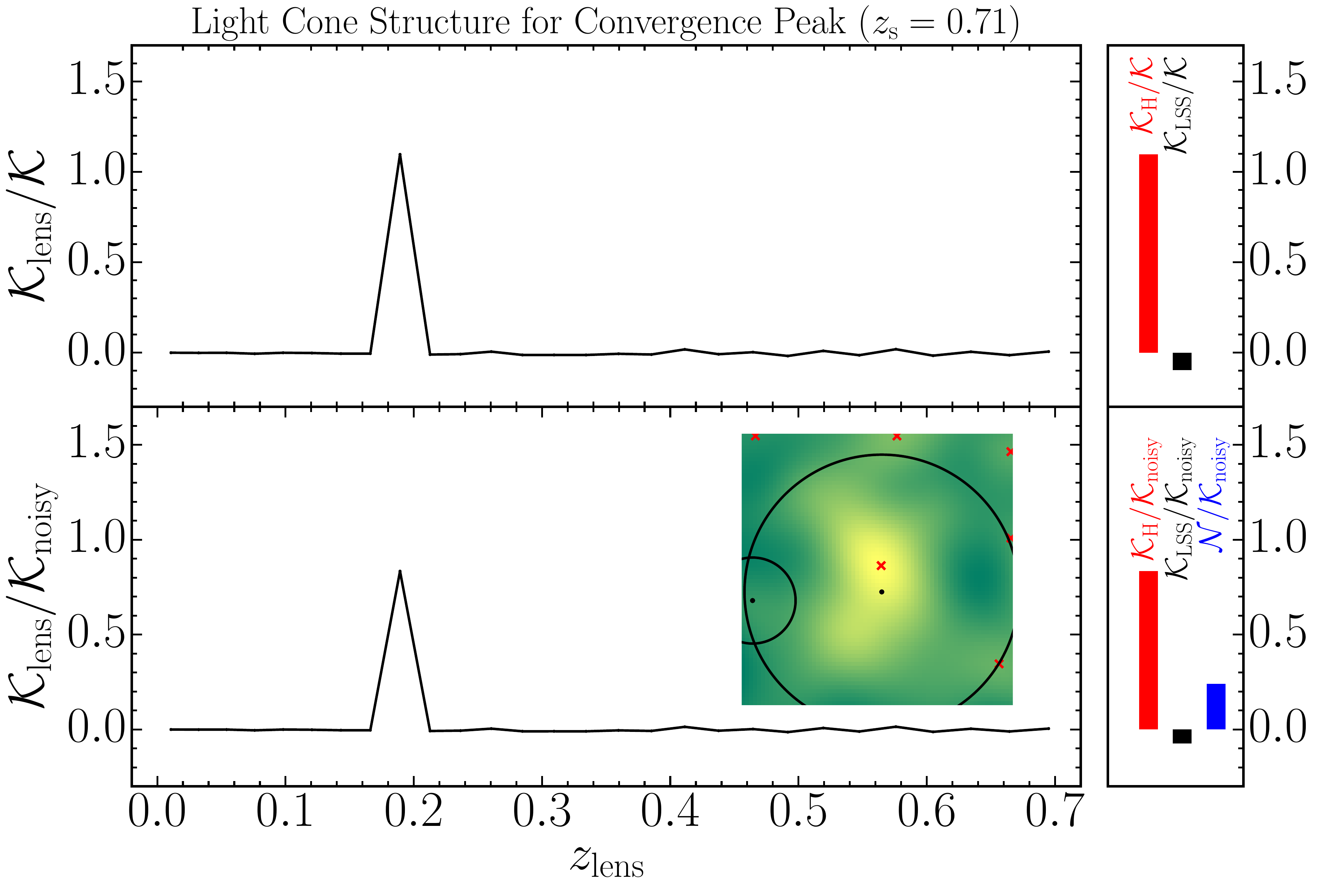}\includegraphics[width=9cm]{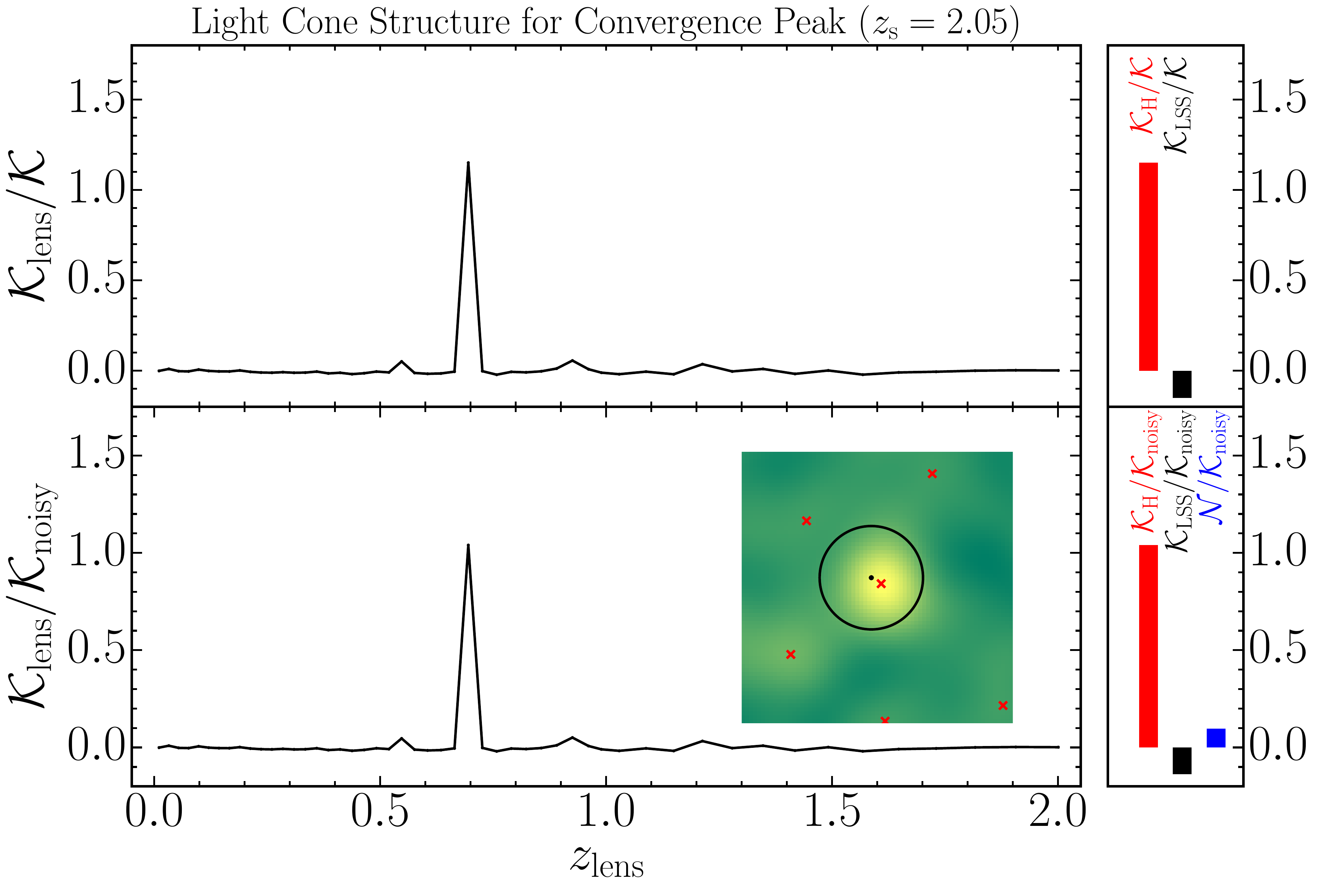}
\caption{The light cone structures of two high peaks for $z_{\mathrm{s}}=0.71$ (left)
and 2.05 (right), respectively. The two noisy peaks with signal-to-noise ratio $\nu_0=6.244$ (left) and $\nu_0=10$ (right). The upper panels show the contribution of each
lens plane to the peak in noise free convergence map, i.e., 
the ratio  $\cal K_{\mathrm{lens}}/\cal K$ where $\cal K=\cal K_{\mathrm{H}} + \cal K_{\mathrm{LSS}}$. The contribution of the dominant halo is labelled as $\cal K_{\mathrm{H}}/\cal K$
as red stick and the LSS projection is labelled as $\cal K_{\mathrm{LSS}}/\cal K$
as black stick. The bottom panels are 
for the cases with shape noise added
($n_{g}=10\ \mathrm{arcmin^{-2}}$
for left and $n_{g}=20\ \mathrm{arcmin^{-2}}$ for right. The smoothing
scales are both $\theta_{G}=2.0\ \mathrm{arcmin.}$) where $\cal K_{\mathrm{nosiy}}=\cal K_{\mathrm{H}} + \cal K_{\mathrm{LSS}}+\cal N$.  The target peak
is in the center of the corresponding stamp, labelled as red cross and the halos ($M\geqslant10^{14.0} h^{-1} M_{\odot}$)
are shown as the black circles with the sizes of their projected angular virial radii.
The contribution of the shape noise $\cal K_{\mathrm{N}}/\cal K_{\mathrm{noisy}}$ is shown as the blue stick.
\label{fig:LOShigh}}
\end{figure*}

In Fig.\ref{fig:LOShigh}, we zoom in two high peaks from our ray-tracing
simulations to be described in detail in \S3. The horizontal axes
are for the redshift of the lens planes, and the vertical axes show
the relative contribution of each lens plane to the final peak signal
${\cal K}_{\mathrm{peak}}$. The upper panels show the noiseless cases
from ray-tracing simulations, and the lower panels are for the cases
adding the shape noise from intrinsic ellipticities of source galaxies.
The left ones are for a peak with the source galaxies at $z_{\mathrm{s}}=0.71$,
and $n_{g}=10\,\mathrm{arcmin^{-2}}$. The insert in the lower panel
shows the zoom-in local image of the noisy peak. Here we apply a Gaussian
smoothing with the window function 
\begin{equation}
W_{\theta_{G}}(\boldsymbol{\theta})=\frac{1}{\pi\theta_{G}^{2}}\exp\bigg(-\frac{|\boldsymbol{\theta}|^{2}}{\theta_{G}^{2}}\bigg).\label{gwindow}
\end{equation}
We take $\theta_{G}=2.0\:\mathrm{arcmin}$ in this paper. For this
peak, ${\cal {K}}_{\mathrm{peak}} \approx0.0835$ for the noiseless case
(upper), and ${\cal {K}}_{\mathrm{peak}}\approx0.114$ for the noisy case
(lower). It is seen clearly that the lens plane at $z\approx0.18$
contributes dominantly to the peak signal. Further examination finds
that there is a massive halo located there. In the noiseless case,
the LSS effect from other lens planes only accounts for less than
10\% (negative) of the peak signal as indicated by the black bar.
In the noisy case, the dominant halo contributes $\sim80\%$ of the
peak signal. The shape noise as indicated by the blue bar contributes
$\sim25\%$ , and LSS from other planes contributes $\sim-5\%$.
The right panels are for the case with $z_{\mathrm{s}}=2.05$ and $n_{g}=20\,\mathrm{arcmin^{-2}}$.
Here the dominant halo is at $z\approx0.7$. In this case, the LSS
effect increases to $\sim-15\%$ because of the increase of $z_{\mathrm{s}}$.
The shape noise contribution is $\sim10\%$.

%F1

These two examples show that for high peaks, the halo approach to
model the WL peak abundances is a physically viable approach. The
shape noise and the LSS projection effect can be regarded as perturbations to
the signal from the dominant halo. For relatively shallow surveys,
the shape noise is much larger than the LSS effect, and we expect
the good performance of the \citetalias{Fan2010} model that takes into account the
shape noise but without including the LSS effect. For deep surveys,
however, the two perturbations are comparable, and the stochastic
LSS effect cannot be neglected.

\begin{figure}
\begin{centering}
\includegraphics[width=9cm]{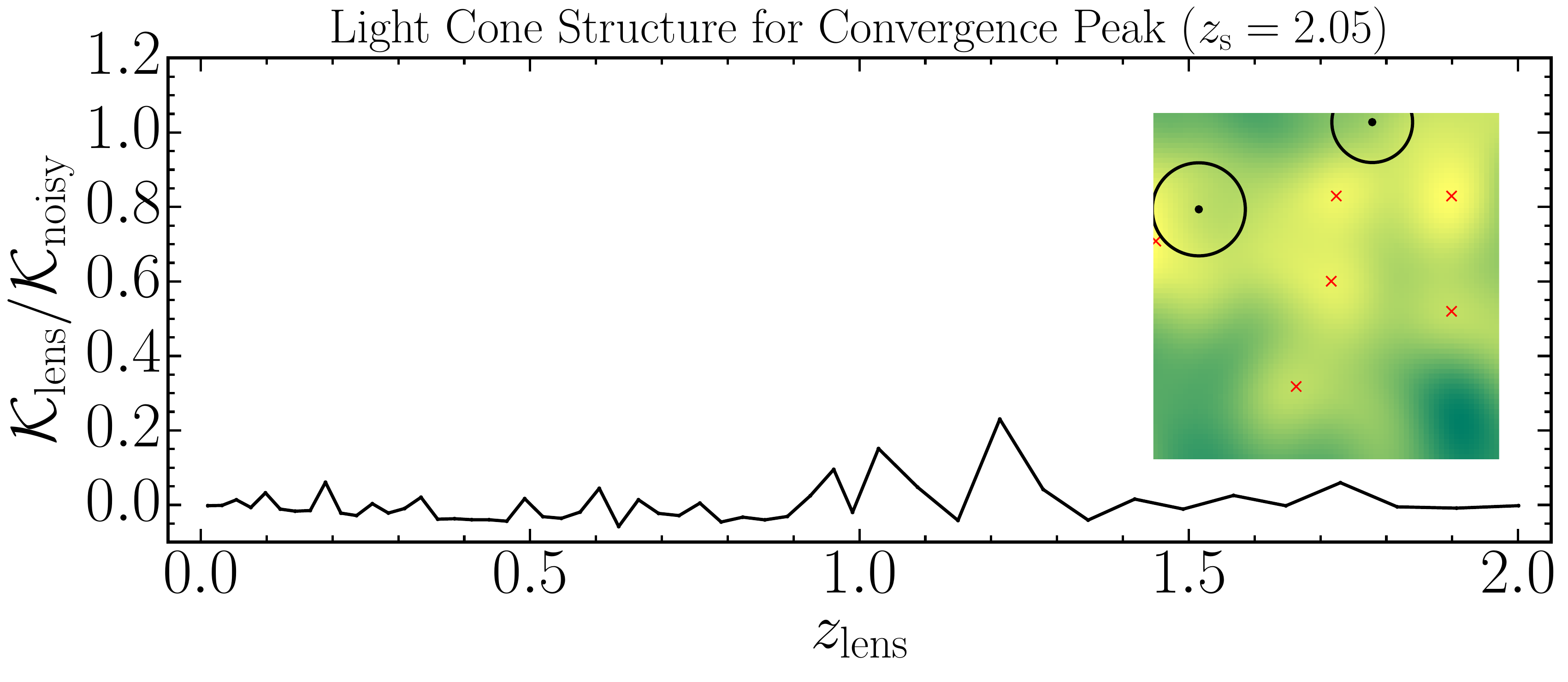} 
\par\end{centering}
\caption{The light cone structure of a  medium peak with the shape noise included ($z_{\mathrm{s}}=2.05$).
\label{fig:LOSsmallpk}}
\end{figure}

As a comparison, we show in Fig.\ref{fig:LOSsmallpk} a low peak
with $z_{\mathrm{s}}=2.05$ and $n_{g}=20\,\mathrm{arcmin}^{-2}$. The peak
signal is ${\cal {K}}_{\mathrm{peak}}\approx0.0422$. %($\nu \approx 3.3$) . 
It is seen that the signals
are from the cumulative effect of the line-of-sight mass distribution
and no dominant halo contribution can be found. For these peaks, different
modelling methodology other than the halo approach is needed.

%F2

In this paper, we focus on high peaks, and present our model for high
peak abundances including LSS projection effects. Similar to \citetalias{Fan2010},
we assume that the signal of a true high peak is mainly from a single
massive halo. The shape noise contributes a  random component
to the reconstructed convergence field. For stochastic LSS, from Fig.\ref{fig:LOShigh},
we see that they add onto the final peak signal in a zigzag way leading
to a random perturbation statistically. Thus it is appropriate to
model the LSS projection effect also as a random field. Therefore
in our model, the convergence field in a halo region can be written
as 
\begin{equation}
\mathcal{K}=\mathcal{K}_{\mathrm{H}}+\mathcal{K}_{\mathrm{LSS}}+\mathcal{N},\label{kall}
\end{equation}
where $\mathcal{K}_\mathrm{H}$ is the contribution from the halo convolved
with a window function corresponding to the smoothing operation made
in the convergence reconstruction. It is regarded as a known quantity
given the density profile of the halo. The smoothed shape noise field
$\mathcal{N}$ is assumed to be Gaussian due to the central limit
theorem \citep[e.g.,][]{Waerbeke2000}. 
We note that in general, the overall convergence field from
simulations show non-Gaussianity due to the nonlinearity of structure
formation. In our consideration here, however, $\mathcal{K}_\mathrm{LSS}$
is the stochastic LSS contribution excluding the massive halo part which is already
explicitly split out as $\mathcal{K}_\mathrm{H}$. It can be more Gaussian
than the overall convergence field. Also $\mathcal{K}_\mathrm{LSS}$ is from
small additive contributions from different lens planes in the high
peak case and $|\mathcal{K}_\mathrm{LSS}| \ll|\mathcal{K}_\mathrm{H}|$. We therefore,
as an approximation, assume that $\mathcal{K}_\mathrm{LSS}$ is also a Gaussian
random field. Its validity will be extensively tested in \S3 by comparing
the model predictions for high peak counts with the results from simulations.

Similar consideration was mentioned in \citet{Shirasaki2015} but
without really calculating the LSS contribution. Also, they only concentrate
on the influence of the random field on the central peak signals of
halos. In our modelling here, we take into account specifically the
stochastic LSS, and calculate the total peak counts, including both the
central ones from massive halos and the peaks from the random field
$\mathcal{K}_\mathrm{LSS}+\mathcal{N}$ inside halo regions as well as outside
halo regions. In other words, to apply our model for cosmological
studies, we can simply use all the high peaks identified from convergence
maps without the need to go through additional analyses to locate
true halo-associated peaks.

With the Gaussian assumptions for the two random fields, the total
field $\mathcal{K}$ in Eq.(\ref{kall}) is also a Gaussian random
field. More specifically, it is the Gaussian random field $\mathcal{K}_\mathrm{LSS}+\mathcal{N}$
modulated by the known halo contribution $\mathcal{K}_\mathrm{H}$. Following
the same procedures shown in \citetalias{Fan2010}, we can then calculate the number
of peaks in a halo region. Two features need to be addressed. First,
the original peak signal from the halo is affected by the existence
of the two random fields, which not only generates scatters, but also
leads to a positive shift for the signal (\citetalias{Fan2010}; \citealp{Shirasaki2015}).
Secondly, the height distribution of peaks generated purely by the
stochastic part $\mathcal{K}_\mathrm{LSS}+\mathcal{N}$ is modulated by the halo
convergence profile $\mathcal{K}_\mathrm{H}$. With the halo mass function,
we can then compute statistically the number of peaks per unit area
in regions occupied by massive halos. For peaks outside the halo regions,
we can calculate the peak abundances simply from the Gaussian field
$\mathcal{K}_\mathrm{LSS}+\mathcal{N}$.

In formulae, for high peak abundances, we have (\citetalias{Fan2010}) 
\begin{equation}
n_{\mathrm{peak}}(\nu)\mathrm{d}\nu=n_{\mathrm{peak}}^{\mathrm{c}}(\nu)\mathrm{d}\nu+n_{\mathrm{peak}}^{\mathrm{n}}(\nu)\mathrm{d}\nu,\label{npksum}
\end{equation}
where $\nu=\mathcal{K}/\sigma_{0}$ with $\sigma_{0}^{2}=\sigma_\mathrm{LSS,0}^{2}+\sigma_\mathrm{N,0}^{2}$
being the total variance of the field $\mathcal{K}_\mathrm{LSS}+\mathcal{N}$,
$n_{\mathrm{peak}}^{\mathrm{c}}(\nu)$ and $n_{\mathrm{peak}}^{\mathrm{n}}(\nu)$ are,
respectively, the number density of peaks per unit $\nu$ centered
at $\nu$ in and outside halo regions.

We emphasize that our model is applicable for high peaks in which
the signals of true peaks are dominated by single massive halos (see
Fig.\ref{fig:LOShigh}). Thus we consider halos with mass $M\geqslant M_{*}$
as major  contributors to the halo regions. Simulation analyses show that $M_{*} \sim 10^{14} h^{-1}{M}_{\odot}$
is a proper choice. \citep{x2018MNRAS.474..712M}. The rest from smaller halos and the correlations
between halos is included in the stochastic LSS part. Then for $n_{\mathrm{peak}}^{\mathrm{c}}(\nu)$,
we have

\begin{multline}
n_{\mathrm{peak}}^{\mathrm{c}}(\nu)=\int\mathrm{d}z\:\dfrac{\mathrm{d}V(z)}{\mathrm{d}z\mathrm{d\Omega}}\int_{M_{*}}^{\infty}\mathrm{d}M\:n(M,z) \\ \times \int_{0}^{\theta_{\mathrm{vir}}}\mathrm{d}\theta(2\pi\theta)\hat{n}^{\mathrm{c}}_{\mathrm{peak}}(\nu,M,z,\theta),\label{eq:npk_c-1}
\end{multline}
where $n(M,z)$ is the comoving halo mass function, $\hat{n}_{\mathrm{peak}}^{\mathrm{c}}(\nu,M,z,\theta)$
is the number density of peaks at $\theta$, and $\theta_{\mathrm{vir}}=R_{\mathrm{vir}}(M,z)/D_{\mathrm{A}}(z)$
is the angular virial radius of a halo with mass $M$ at redshift
$z$. Here $R_{\mathrm{vir}}$ and $D_{\mathrm{A}}$ are the virial
radius of the halo and the angular diameter distance to the halo,
respectively.

Considering the Gaussian random field $\mathcal{K}_\mathrm{LSS}+\mathcal{N}$
modulated by the halo term $\mathcal{K}_\mathrm{H}$, following the calculations
in \citetalias{Fan2010}, we have

\begin{eqnarray}
 &  & \hat{n}_{\mathrm{peak}}^{\mathrm{c}}(\nu,M,z,\theta)=\exp\bigg[-\frac{(\mathcal{K}_\mathrm{H}^{1})^{2}+(\mathcal{K}_\mathrm{H}^{2})^{2}}{\sigma_{1}^{2}}\bigg]\nonumber \\
 &  & \times\bigg[\frac{1}{2\pi\theta_{*}^{2}}\frac{1}{(2\pi)^{1/2}}\bigg]\exp\bigg[-\frac{1}{2}\bigg(\nu-\frac{\mathcal{K}_\mathrm{H}}{\sigma_{0}}\bigg)^{2}\bigg]\nonumber \\
 &  & \times\int_{0}^{\infty}dx\bigg\{\frac{1}{[2\pi(1-\gamma^{2})]^{1/2}}\nonumber \\
 &  & \times\exp\bigg[-\frac{[{x+(\mathcal{K}_\mathrm{H}^{11}+\mathcal{K}_\mathrm{H}^{22})/\sigma_{2}-\gamma(\nu-\mathcal{K}_\mathrm{H}/\sigma_{0})}]^{2}}{2(1-\gamma^{2})}\bigg]\nonumber \\
 &  & \times F(x)\bigg\}\label{nchat}
\end{eqnarray}
and 
\begin{eqnarray}
 &  & F(x)=\exp\bigg[-\frac{(\mathcal{K}_\mathrm{H}^{11}-\mathcal{K}_\mathrm{H}^{22})^{2}}{\sigma_{2}^{2}}\bigg]\times\nonumber \\
 &  & \int_{0}^{1/2} \mathrm{d}e \hbox{}8(x^{2}e)x^{2}(1-4e^{2})\exp(-4x^{2}e^{2})\times\nonumber \\
 &  & \int_{0}^{\pi}\frac{d\psi}{\pi}\hbox{}\exp\bigg[-4xe\cos(2\psi)\frac{(\mathcal{K}_\mathrm{H}^{11}-\mathcal{K}_\mathrm{H}^{22})}{\sigma_{2}}\bigg].\nonumber \\
\label{fx}
\end{eqnarray}
Here $\theta_{*}^{2}=2\sigma_{1}^{2}/\sigma_{2}^{2}$, $\gamma=\sigma_{1}^{2}/(\sigma_{0}\sigma_{2})$,
$\mathcal{K}_\mathrm{H}^{i}=\partial_{i}\mathcal{K}_\mathrm{H}$, and $\mathcal{K}_\mathrm{H}^{ij}=\partial_{ij}\mathcal{K}_\mathrm{H}$.
Different from that in \citetalias{Fan2010}, here the quantities $\sigma_{i}^{2}$
($i=0,1,2$) are, respectively, the moments of the total random field
$\mathcal{K}_\mathrm{LSS}+\mathcal{N}$ and its first and second derivatives.
Specifically, $\sigma_{i}^{2}=\sigma_{\mathrm{LSS},i}^{2}+\sigma_{\mathrm{N},i}^{2}$.

%F3
\begin{figure}
\begin{centering}
\includegraphics[scale=0.35]{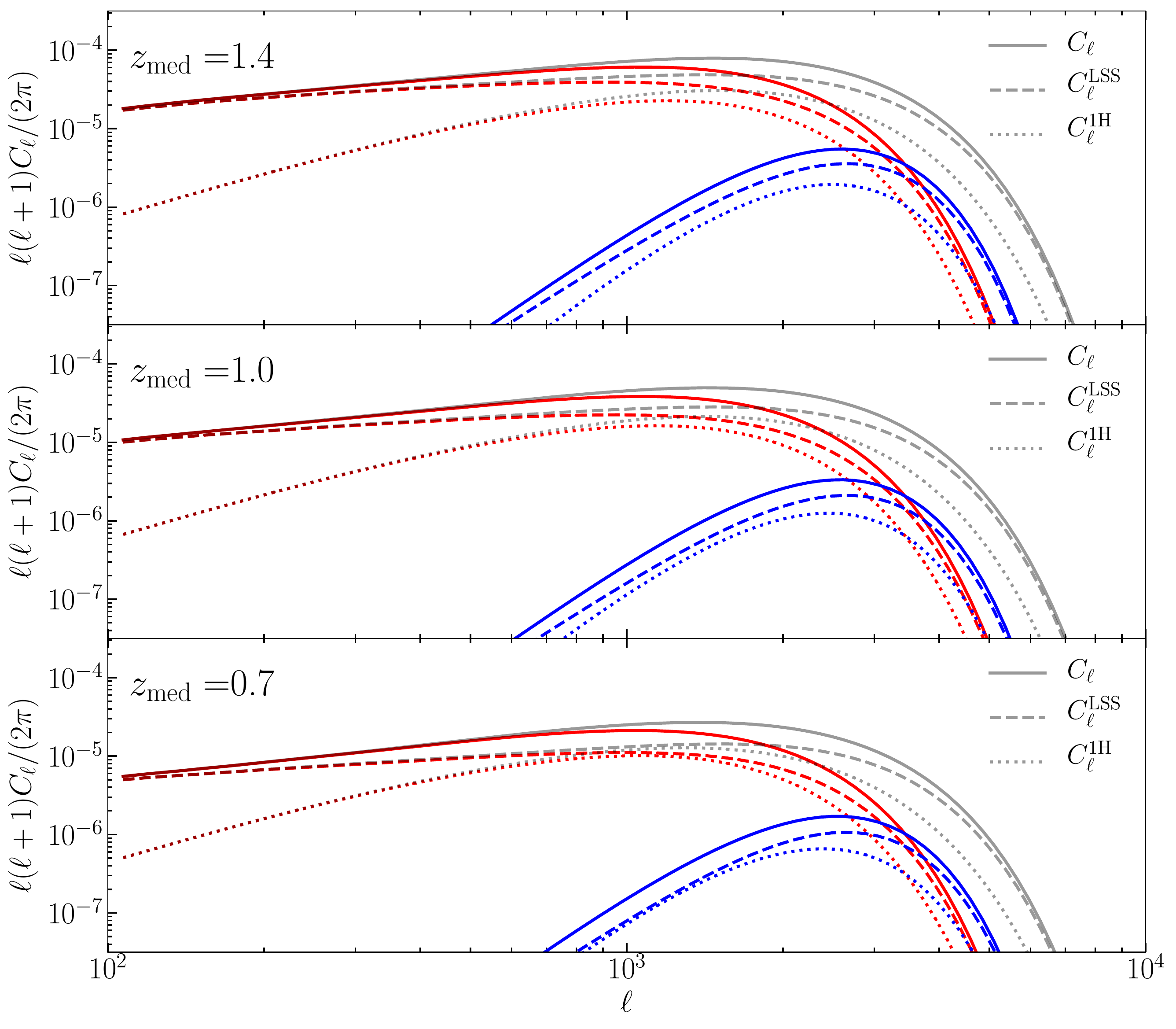}
\par\end{centering}
\caption{\label{fig:GUcl}
Weak lensing power spectra (gray), and the ones under Guassian smoothing (red) and the U filtering (blue).The dashed and the dotted lines are the contributions of the LSS and the one-halo term from halos with $M\ge M_*$.
The smoothing scale $\theta_{G}=\theta_{U}=2.0$ arcmin is applied. The redshifts are
$z_{{\rm med}}=1.4,1.0$ and 0.7, respectively, from top to bottom.
The cosmological parameters are $(\Omega_{m},\Omega_{\Lambda},\Omega_{b},h,n_{s},\sigma_{8})=(0.28,0.72,0.046,0.7,0.96,0.82)$ 
}
\end{figure}

For the number density of peaks contributed by those outside halo
regions, $n_{\mathrm{peak}}^{\mathrm{n}}(\nu)$, we have 
\begin{multline}
n_{\mathrm{peak}}^{\mathrm{n}}(\nu)= \frac{1}{\mathrm{d}\Omega} \times \\ \left\{ n_{\mathrm{ran}}(\nu)\left[\mathrm{d\Omega-\int\mathrm{d}}z\:\frac{\mathrm{d}V(z)}{\mathrm{d}z} \int_{M_{*}}^{\infty}\mathrm{d}M\:n(M,z)(\pi\theta_{\mathrm{vir}}^{2})\right]\right\} ,\label{eq:npk_n-1}
\end{multline}
where $n_{\mathrm{ran}}(\nu)$ is the number density of peaks from
the random field $\mathcal{K}_\mathrm{LSS}+\mathcal{N}$ without halo modulations,
and it can be calculated from Eq.(\ref{nchat}) by setting the halo
related quantities to be zero.

%F4
\begin{figure*}
\begin{centering}
\includegraphics[width=9cm]{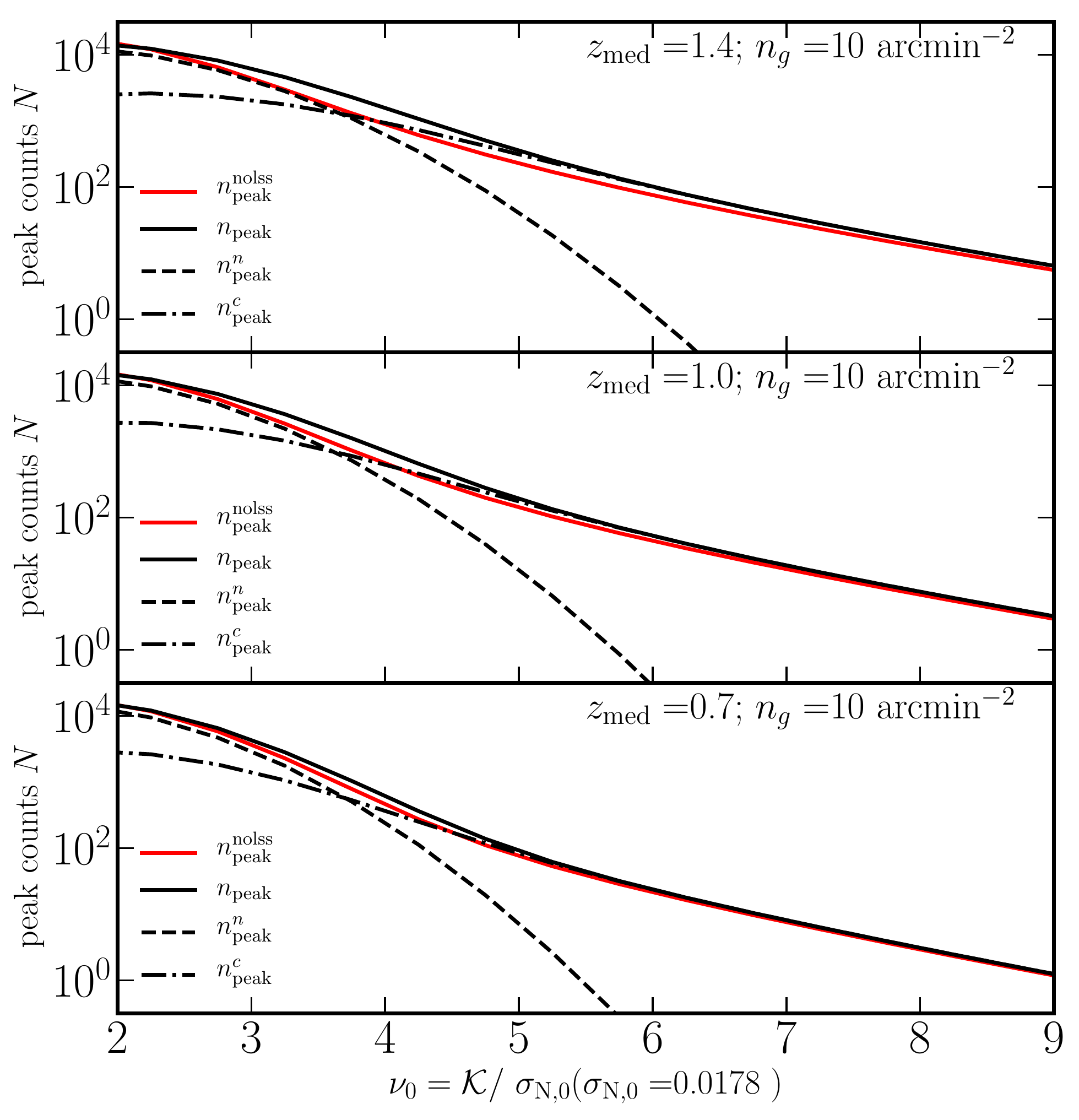}\includegraphics[width=9cm]{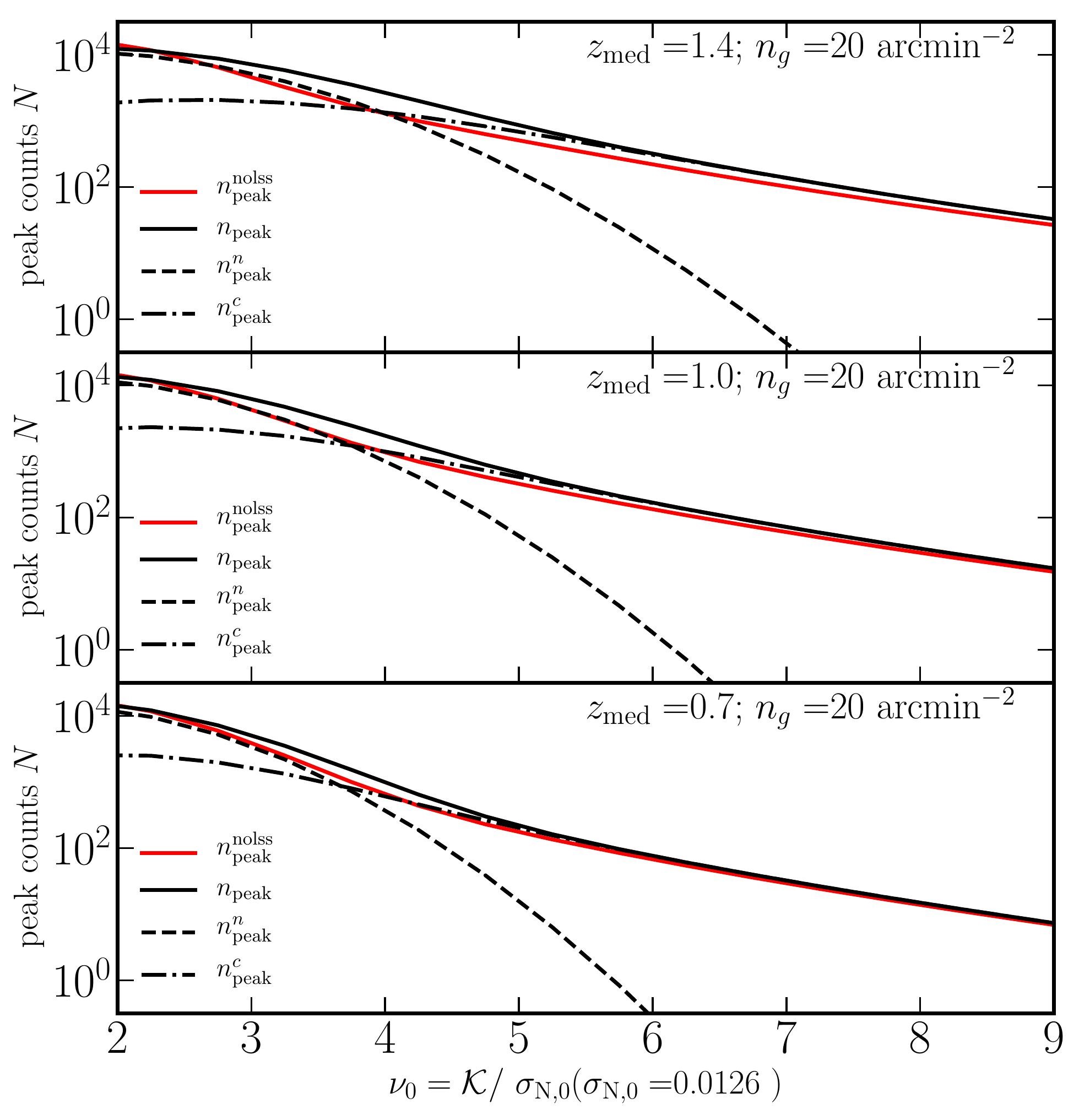} 
\par\end{centering}
\caption{\label{fig:oldvsnew}Model predictions from \citetalias{Fan2010} (red lines) and  from our model  considering LSS
projection effect (black lines) with different $z_{\mathrm{med}}$. From top to bottom  $z_{\mathrm{med}}=1.4,1.0$
    and 0.7, respectively and the sky area is $\approx 1080\, \mathrm{deg}^2$. The cosmological parameters are $(\Omega_{m},\Omega_{\Lambda},\Omega_{b},h,n_{s},\sigma_{8})=(0.28,0.72,0.046,0.7,0.96,0.82)$ . The components $n_{{\rm peak}}^{n}$and $n_{{\rm peak}}^{c}$ of total
$n_{{\rm peak}}$ with LSS projection effect included are shown in dashed lines and dashed-dotted lines, respectively.  The smoothing scale is $\theta_{G}=2.0$ arcmin, the source number density $n_{g}=10\,{\rm arcmin^{-2}}$(left panel) and
$20\ \mathrm{arcmin}^{-2}$ (right panel)}
\end{figure*}

From the above, we see that the stochastic LSS effects occur specifically in
quantities of $\sigma_{i}^{2}=\sigma_{\mathrm{LSS},i}^{2}+\sigma_{\mathrm{N},i}^{2}$.
For the shape noise part, we have (e.g. \citealt{Waerbeke2000}) 
\begin{equation}
\sigma_{\mathrm{N},i}^{2}=\int_{0}^{\infty}\frac{\ell\mathrm{d}\ell}{2\pi}\ell^{2i}C_{\ell}^{\mathrm{N}},\label{noisep}
\end{equation}
where $C_{\ell}^{\mathrm{N}}$ is power spectrum of the smoothed noise field
$\mathcal{N}$. For the Gaussian smoothing, we have
\begin{equation}
\sigma_{\mathrm{N},0}^{2}=\cfrac{\sigma_{\epsilon}^{2}}{4\pi n_{g}\theta_{G}^{2}},
\end{equation}
where $\sigma_{\epsilon}$ is the rms amplitude of the intrinsic ellipticities.
We further have $\sigma_{\mathrm{N},0}:\sigma_{\mathrm{N},1}:\sigma_{\mathrm{N},2}=1:\sqrt{2}/\theta_{G}:2\sqrt{2}/\theta_{G}^{2}$.

For $\sigma_{\mathrm{LSS},i}$, they are physical quantities and need to be
computed in a cosmology-dependent way. In other words, $\sigma_{\mathrm{LSS},i}$
also contributes to the cosmological information embedded in WL peak
counts. Given a power spectrum for LSS convergence field, $C_{\ell}^{\mathrm{LSS}}$, we have
\begin{equation}
\sigma_{\mathrm{LSS},i}^{2}=\int_{0}^{\infty}\frac{\ell\mathrm{d}\ell}{2\pi}\ell^{2i}C_{\ell}^{\mathrm{LSS}}.\label{lssp}
\end{equation}

To calculate $C_{\ell}^{\mathrm{LSS}}$, we adopt the following approach.
From Eq.(\ref{eq:Limb}), the WL power spectrum $C_{\ell}$ can
be obtained from the integration of the weighted 3-D nonlinear power
spectrum $P_{\delta}$. For $P_{\delta}$, it can be computed using
the simulation-calibrated halo model \citep{Takahashi2012} and has
been included in different numerical packages, such as \texttt{CAMB}
\citep{Lewis2000}. In the language of halo model, the overall $P_{\delta}$
consists of contributions from one-halo term of all halos and the
two-halo term considering the correlations between halos. In our model
here, halos with $M\geqslant M_{*}$ have been separated out as $\mathcal{K}_\mathrm{H}$.
Thus to compute the left-over stochastic LSS effects, we subtract the one-halo
term from halos with $M\geqslant M_{*}$ from the overall power spectrum:
\begin{equation}
P_{\delta}^{\mathrm{LSS}}[k,\chi(z)]=P_{\delta}[k,\chi(z)]-  P_{\delta}^{\mathrm{1H}} \Bigm|  _{M \geqslant M_{*}}[k,\chi(z)].\label{plss}
\end{equation}
It is seen that $P_{\delta}^{\mathrm{LSS}}$ contains the one-halo term from halos
with $M<M_{*}$ and the two-halo term between all the halos including
the ones with $M\geqslant M_{*}$. For the one-halo term $P_{\delta}^{\mathrm{1H}} \Bigm| _{M\geqslant M_{*}}$,
we have \citep[e.g.][]{Cooray2002} 
\begin{equation}
P_{\delta}^{\mathrm{1H}} \Bigm| _{M\geqslant M_{*}}[k,\chi(z)]=\frac{4\pi}{\bar{\rho}^{2}}\int_{M_{*}}^{\infty}\mathrm{d}M\,M^{2}W^{2}(k,M)n(M,z),\label{eq:p-1h}
\end{equation}
where $\bar{\rho}$ is the mean matter density of the universe, $W(k,M)$
is the Hankel transformation of the spherically symmetric halo density
profile $\rho(r,M)$, given by 
\begin{equation}
W(k,M)=\frac{1}{M}\int_{0}^{R_{\mathrm{vir}}}\mathrm{d}r\,\frac{\sin(kr)}{kr}4\pi r^{2}\rho(r,M).
\end{equation}

From $P_{\delta}^{\mathrm{LSS}}$,
we can obtain $C_{\ell}^{\mathrm{LSS}}$ by 
\begin{equation}
C_{\ell}^{\mathrm{LSS}}=\frac{9H_{0}^{4}\Omega_{m}^{2}}{4}\int_{0}^{\chi_{H}}\mathrm{d}\chi'\frac{w^{2}(\chi')}{a^{2}(\chi')}P_{\delta}^{\mathrm{LSS}}\Bigl(\frac{\ell}{f_{K}(\chi')},\chi'\Bigr),\label{cllss}
\end{equation}
and subsequently $\sigma_{\mathrm{LSS},i}^{2}$ by Eq.(\ref{lssp}).

For the calculations of one-halo term in Eq.(\ref{eq:p-1h}), we take
the Navarro-Frenk-White (NFW) halo density profile given by \citep{Navarro1996,Navarro1997}
\begin{equation}
\rho(r)=\frac{\rho_{\mathrm{s}}}{(r/r_{\mathrm{s}})(1+r/r_{\mathrm{s}})^{2}},\label{eq:nfw}
\end{equation}
where $\rho_{\mathrm{s}}$ and $r_{\mathrm{s}}$ are the characteristic density and
scale of a halo, respectively. The scale $r_{\mathrm{s}}$ reflects the compactness
of a halo, and is often given through the concentration parameter
$c_{\mathrm{th}}=R_{\mathrm{th}}/r_{\mathrm{s}}$ with $R_{\mathrm{th}}$ being the radius inside which
the average density of a halo is $\Delta_{\mathrm{th}}$ times the cosmic density.
Here we adopt the virial radius $R_{\mathrm{vir}}$, and use the concentration-mass
relation from \citet{Duffy2008} with

\begin{equation}
c_{\mathrm{vir}}(M,z)=5.72\Bigl(\frac{M}{10^{14}h^{-1}M_{\odot}}\Bigr)^{-0.081}(1+z)^{-0.71}.\label{eq:cm}
\end{equation}

For the halo mass function $n(M,z)$, we use the one given in \citet{Watson2013},
an empirical fitting formula derived from $N$-body simulations.

In Fig.\ref{fig:GUcl}, we show $C_{\ell}^{\mathrm{LSS}}$ together with
the overall  $C_{\ell}$ and the massive one-halo term $C_{\ell}^{{\rm 1H}}$
under different source galaxy distributions $p(z)$, for which we adopt the
following form 
\begin{equation}
p(z)\propto z^{2}\exp\left[-\left(1.414\frac{z}{z_{\mathrm{med}}}\right)^{1.5}\right],\label{pz}
\end{equation}
where $z_{\mathrm{med}}$ is the median redshift.  
In the plots, we also show the power spectra of aperture mass under the U filtering (blue) to be disscussed in Sec.\ref{sec:Uf}. Here we focus on the Gaussian filter case (red).
We see that on large
scales, $C_{\ell}^{\mathrm{LSS}}$ is very close to $C_{\ell}$. 
On small scales
with $\ell\sim2000$, the LSS random field $C_{\ell}^{\mathrm{LSS}}$ is small than
the overall convergence field $C_{\ell}$ due to the exclusion of the one-halo term
from halos with $M\geqslant10^{14.0} h^{-1} \mathrm{M}_\odot$

With Eq.(\ref{npksum}) to Eq.(\ref{cllss}), the number density of
high peaks taking into account the LSS projection effects can be computed.
In Fig.\ref{fig:oldvsnew}, we show the results from this model (solid lines) and
the ones from \citetalias{Fan2010} without the LSS (red lines). 
For the model with LSS effects, we also show the contributions of peaks in ($n_\mathrm{peak}^c$, dashed-dotted) and outside ($n_\mathrm{peak}^n$, dashed) halo regions.
It is seen that in the considered cases, peaks with $\nu_0 \gtrsim 4$ are dominantly from halo regions. For higher $z_{\rm med}$, such domination shifts a little more toward higher $\nu_0$.

We note that in our model
calculation, we directly obtain peak counts at different $\nu=\cal{K} / \sigma_\mathrm{0}$.
On the other hand, observationally, we can only estimate the shape
noise part $\sigma_\mathrm{N,0}$ by randomly rotating the observed galaxies.
Thus to be consistent with observational analyses, we first make a
binning in terms of $\nu_{0}=\cal K  /\sigma_\mathrm{N,0}$ and then convert
it to the binning in $\nu$ using the ratio of $\sigma_\mathrm{N,0}/\sigma_{0}$
for model calculations. The shown results are the peak counts versus $\nu_{0}$. The corresponding $\cal K$ are also listed in the upper
horizontal axis. It is seen clearly that with the increase of the
median redshift of source galaxies, the LSS projection effects become increasingly
important.

In the next section, we will compare our theoretical results with
those from ray-tracing simulations to validate the model performance.

\section{Simulation Tests}
In this section, we test our model performance using ray-tracing simulations.
We describe the simulations and the mock WL data generation with respect
to different source galaxy distributions in \S3.1, and present the
comparison results in detail in \S3.2.

\subsection{WL Simulations}

We carry out ray-tracing simulations up to $z=3.0$ based on large
sets of $N$-body simulations. The simulation setting is the same
as that in {\color{cyan}X. K. }\citet{Liu2015}, but with the number of simulations
doubled. The fiducial cosmology is the flat $\Lambda$CDM model with
the parameters of $\Omega_{m}$,
dark energy density $\Omega_{\Lambda}$, baryonic matter density $\Omega_{b}$,
Hubble constant $h$, the power index of initial matter density perturbation
power spectrum $n_{s}$, and $\sigma_{8}$ set to be $(\Omega_{m},\Omega_{\Lambda},\Omega_{b},h,n_{s},\sigma_{8})=(0.28,0.72,0.046,0.7,0.96,0.82).$

\begin{figure*}
\begin{centering}
\includegraphics[scale=.3]{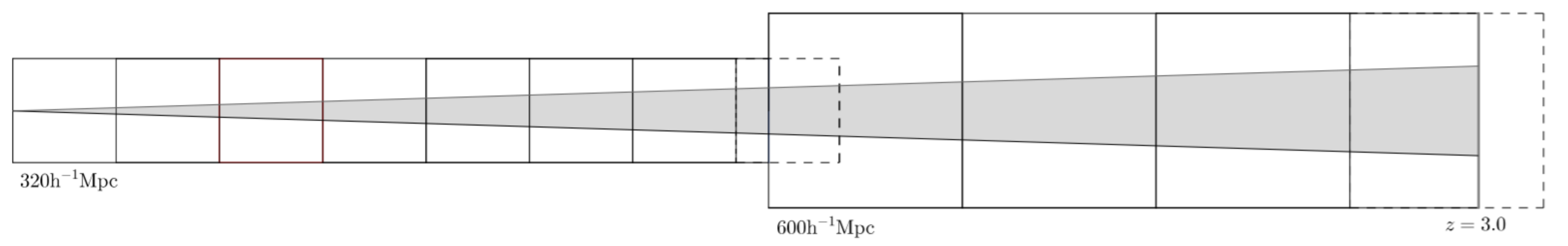} 
\par\end{centering}
\caption{\label{fig:pencil}An illustration of the ray-tracing design from $z=0$ to $z=3$.}
\end{figure*}

For each set of ray-tracing calculations, we use 12 independent $N-$body
simulation boxes to fill up to the region of a comoving distance $\sim4.5\,h^{-1}\mathrm{Gpc}$
to $z=3.0$, as illustrated in Fig.\ref{fig:pencil}. Among them,
eight small boxes each with the size of 320$\,h^{-1} \mathrm{Mpc}$ are padded
between $z=0.0$ and $z=1.0$. In the reshift range of $1.0<z\leqslant3.0$,
we pad four boxes of size $600h^{-1} \mathrm{Mpc}$. The number of particles
of $N-$body simulations
for both small and large boxes is $640^{3}$, and the corresponding
mass resolution is $\sim9.7\times10^{9}h^{-1} M_{\odot}$ and $\sim6.4\times10^{10}h^{-1}M_{\odot}$,
respectively. For each of the boxes, we start at $z=50$ and generate
the initial conditions using \texttt{2LPTic}\footnote{\texttt{\url{http://cosmo.nyu.edu/roman/2LPT/}}}
based on the initial power spectrum from\texttt{ CAMB}\footnote{\texttt{\url{http://camb.info/}}}
(\citealt{Lewis2000}). The simulations are run by \texttt{GADGET-2}\footnote{\url{http://wwwmpa.mpa-garching.mpg.de/gadget/}}
\citep{Springel2005} with the force softening length being $\sim20\,h^{-1}\mathrm{kpc}$.

%F5

In the multi-plane ray-tracing calculations, up to $z=3$, we use
59 lens planes with the corresponding redshifts $z_{l}$ being listed
in Table \ref{tab:zlenslist}. We run the ray-tracing WL simulations
using the same code described in {\color{cyan}X. K. }\citet{Liu2014} in which we deal
with the crossing-boundary problem of halos following the procedures
used in \citet{Hilbert2009}. We then generate $4\times(3.5^{\circ}\times3.5^{\circ}) $
convergence and shear maps at each lens plane, denoted as $\kappa(z_{\mathrm{lens}})$
and $\gamma_{i}(z_{\mathrm{lens}})$, respectively, from a set of 12 $N$-body
simulations. Each of the $3.5^{\circ}\times3.5^{\circ} $ maps is pixelized
into $1024\times1024$ grids with the pixel size of $\sim0.205\,\mathrm{arcmin}$.
We perform in total 24 sets of $N$-body simulations, which give rise
to $24\times4\times(3.5^{\circ}\times3.5^{\circ})=1176\deg^{2}$ of $\kappa(z_{\mathrm{lens}})$
and $\gamma_{i}(z_{\mathrm{lens}})$. From them, we construct the final $\kappa$
(or $\gamma$) maps corresponding to different source galaxy distributions
as follows

\begin{equation}
\kappa_{\mathrm{mock}}(\boldsymbol{\theta})=\sum_{\mathrm{lens}}p(z_{\mathrm{lens}})\kappa(\boldsymbol{\theta};z_{\mathrm{lens}})(z_{\mathrm{lens}+1}-z_{\mathrm{lens}}),
\end{equation}
where $p(z)$ is the normalized source galaxy redshift distribution
given in Eq.(\ref{pz}). For a given $p(z)$, we obtain $24\times4=96$
maps, each with the size of $3.5^{\circ}\times3.5^{\circ}$.

Because we aim at testing our WL high peak model, we concentrate on
convergence maps directly here. We include the shape noise by adding
a Gaussian noise field to the pixels of each of $24\times4=96$ convergence
maps $\kappa_{\mathrm{mock}}(\boldsymbol{\theta})$ with the variance
given by

\begin{equation}
\sigma_{\mathrm{pix}}^{2}=\frac{\sigma_{\epsilon}^{2}}{2n_{g}\theta_{\mathrm{pix}}^{2}},
\end{equation}
where we take $\sigma_{\epsilon}=0.4$ and the pixel size of maps $\theta_{\mathrm{pix}}=0.205\, \mathrm{arcmin}$.
We then apply a Gaussian smoothing given by Eq.(\ref{gwindow}) with
$\theta_{G}=2.0\,\mathrm{arcmin}$ to obtain the final smoothed noisy
convergence maps for peak analyses.

%%%%V

\begin{table}

\caption{Redshifts of the lens planes.\label{tab:zlenslist}}

\begin{centering}
\begin{tabular}{cccccc}
\toprule 
0.0107  & 0.0322  & 0.0540  & 0.0759  & 0.0981  & 0.1205 \tabularnewline
%\midrule 
0.1432  & 0.1661  & 0.1893  & 0.2127  & 0.2364  & 0.2604\tabularnewline
%\midrule 
0.2847  & 0.3094  & 0.3343  & 0.3596  & 0.3853  & 0.4113\tabularnewline
%\midrule 
0.4377  & 0.4645  & 0.4917  & 0.5193  & 0.5474  & 0.5759\tabularnewline
%\midrule 
0.6049  & 0.6344  & 0.6645  & 0.6950  & 0.7261  & 0.7578\tabularnewline
%\midrule 
0.7900  & 0.8229  & 0.8564  & 0.8906  & 0.9254  & 0.9610\tabularnewline
%\midrule 
0.9895  & 1.0289  & 1.0882  & 1.1496  & 1.2131  & 1.2789\tabularnewline
%\midrule 
1.3472  & 1.4180  & 1.4915  & 1.5680  & 1.6475  & 1.7303\tabularnewline
%\midrule 
1.8166  & 1.9066  & 2.0005  & 2.0987  & 2.2013  & 2.3087\tabularnewline
%\midrule 
2.4213  & 2.5393  & 2.6632  & 2.7934  & 2.9296  & \tabularnewline
%\bottomrule
%\toprule
\hline 
\end{tabular}
\par\end{centering}
\end{table}

%%---------

\subsection{Model test}

To analyze the LSS effects and test our model performance, we consider
different survey parameters, including the median redshift $z_{\mathrm{med}}$, 
the number density $n_{g}$ of the source galaxies and the survey
area $S$. These are listed in Table \ref{tab:Mocks}.
%%---------
\begin{table}[h]
\noindent \centering{}\caption{Survey parameters for mocks.\label{tab:Mocks}}
\begin{tabular}{ccccc}
\toprule 
$z_{\mathrm{med}}$  & $n_{g}(\mathrm{arcmin^{-2}})$  & $\theta_{G}(\mathrm{arcmin})$  & $S(\deg^{2})$  & Mock Name\tabularnewline
\hline
 & 10  & 2.0  & 150  & \textbf{S10small}\tabularnewline
\hline 
0.7  & 10  & 2.0  & $\approx1086$  & \textbf{S10}\tabularnewline
\hline
 & 20  & 2.0  & $\approx1086$  & \textbf{S20}\tabularnewline
\hline 
1.0  & 20  & 2.0  & $\approx1086$  & \textbf{M20}\tabularnewline
\hline 
1.4  & 20  & 2.0  & $\approx1086$  & \textbf{D20}\tabularnewline
%\bottomrule
\toprule 

\end{tabular}
\end{table}

From our simulations, for each set of $(z_{\mathrm{med}},n_{g})$,
we generate 96 noiseless convergence maps each with the size of $3.5^{\circ} \times3.5^{\circ}$.
For each map, we then add a Gaussian shape noise field and apply
smoothing as described above. To suppress the fluctuations caused
by a particular realization of the noise field, we perform noise adding
20 times for each map with different seeds. Therefore in total, we
have $20\times96$ maps with the shape noise included for each set
of $(z_{\mathrm{med}},n_{g})$.

We first compare peak counts between model predictions and the simulation
results. We identify a peak in a pixelized convergence map from simulations
if its convergence value is higher than those of its eight neighboring
pixels. Because Fourier transformations are involved in ray-tracing
calculations and in the smoothing operations, there can be boundary
effects in each of the $3.5^{\circ}\times3.5^{\circ}$ convergence maps. To
avoid such a problem, in our peak analyses, we exclude the outermost
20 pixels along each side of the map. The left-over area is $\sim11.31\deg^{2}$
for each map, and the total is $\sim96\times11.31\thickapprox1086\deg^{2}$
for each set of noise field realizations.

For our theoretical model calculations, the quantity $M_{*}$ corresponds
to the lower mass limit of halos above which the halos dominate the
WL high peak signals. We have performed $\chi^{2}$ tests with respect
to the simulated peak counts to find suitable $M_{*}$. We note that
for $z_{\mathrm{med}}=0.7$ and $n_{g}=10\,\mathrm{arcmin}^{-2}$ , the shape
noise is much larger than that of the LSS effects, and the model of
\citetalias{Fan2010} works equally well. In that case using \citetalias{Fan2010}, $M_{*}=10^{13.9}h^{-1}M_{\odot}$
gives results that are in good agreement with those from simulations.
In our current model with the LSS effects, for all the cases including
the one with $z_{\mathrm{med}}=0.7$ and $n_{g}=10\,\mathrm{arcmin}^{-2}$, $M_{*}=10^{14.0}h^{-1}M_{\odot}$ is a proper choice. We comment
that physically, the suitable choice of $M_{*}$ depends on the  
halo mass function used in the model calculations. The specific value of $M_{*}$
may also have a mild cosmology-dependence, which may need to be taken
into account in future for very high precision studies. In this paper,
we do not include this subtle effect. %

%F6
\begin{figure}[h]
\begin{centering}
\includegraphics[width=9cm]{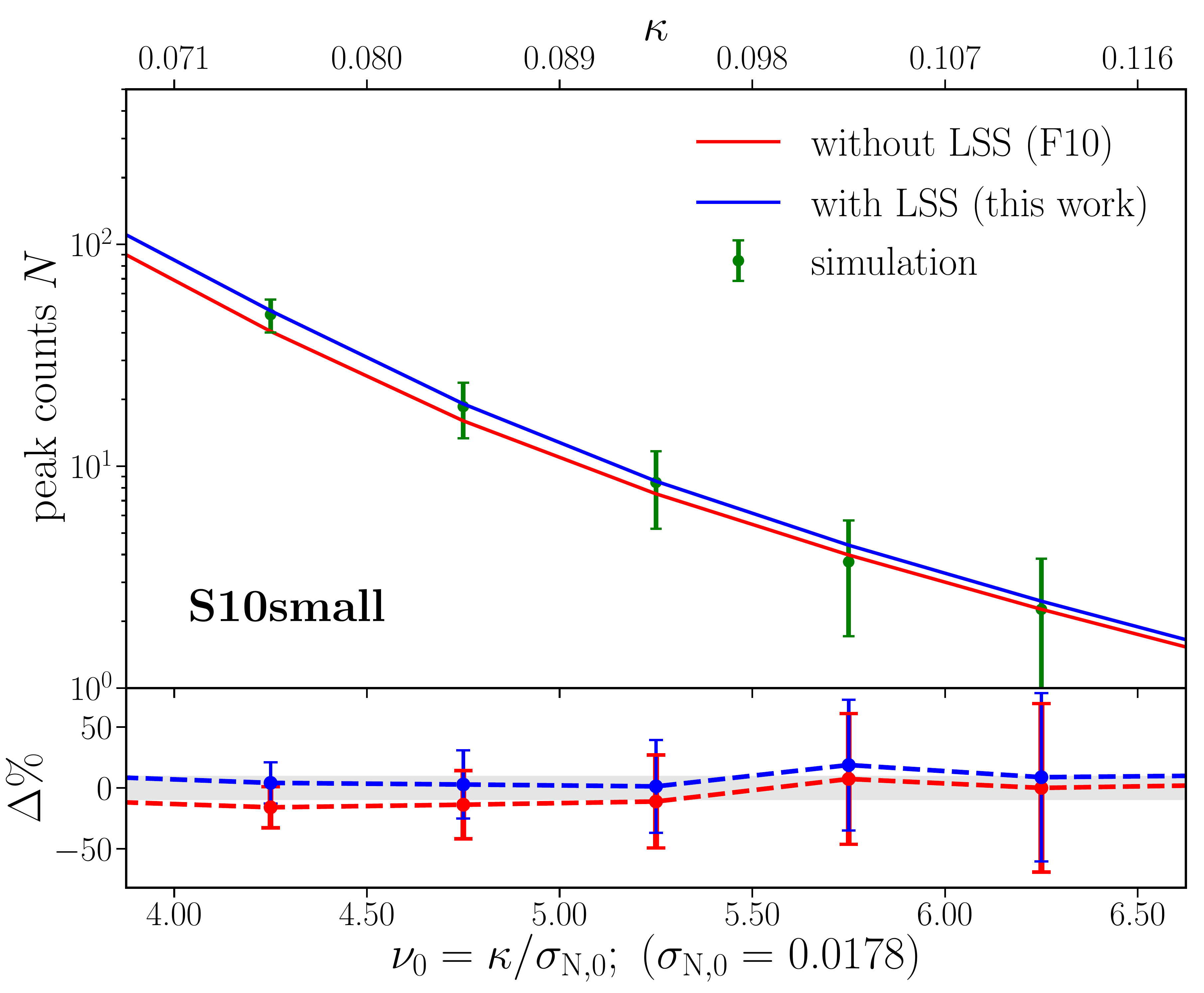} 
\par\end{centering}
\caption{
\label{fig:mock_pre-1}
Model prediction and the peak counts from simulation for \textbf{S10small}. The red line is model prediction
from \citetalias{Fan2010} and the blue line is the  prediction from our model. Green dots
with error bars are peak counts from simulation. The relative  differences  between
simulation results and model predictions are shown in the bottom panel.
}
\end{figure}

\begin{figure*}
\begin{centering}
\includegraphics[width=9cm]{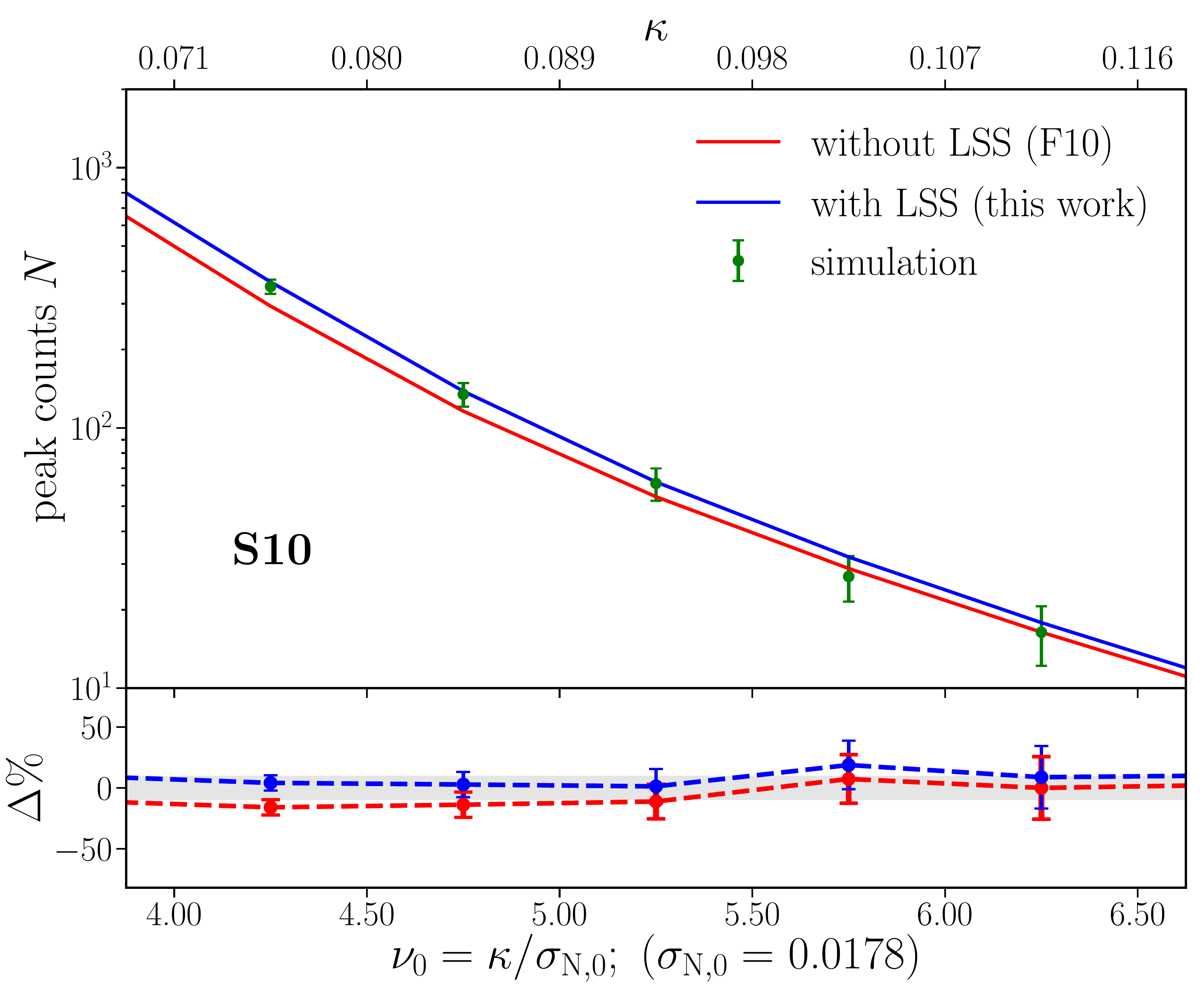}\includegraphics[width=9cm]{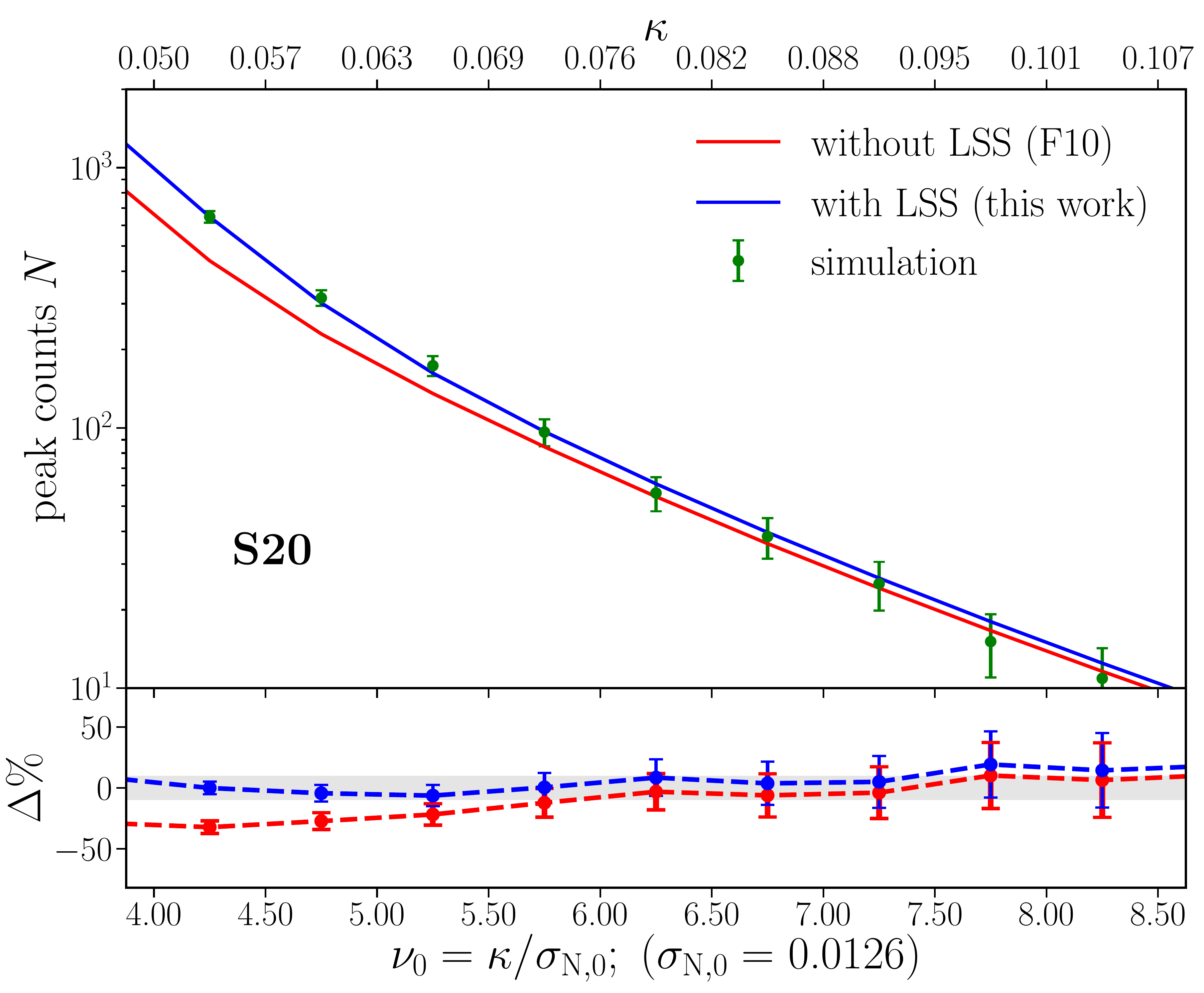} 
\par\end{centering}
\begin{centering}
\includegraphics[width=9cm]{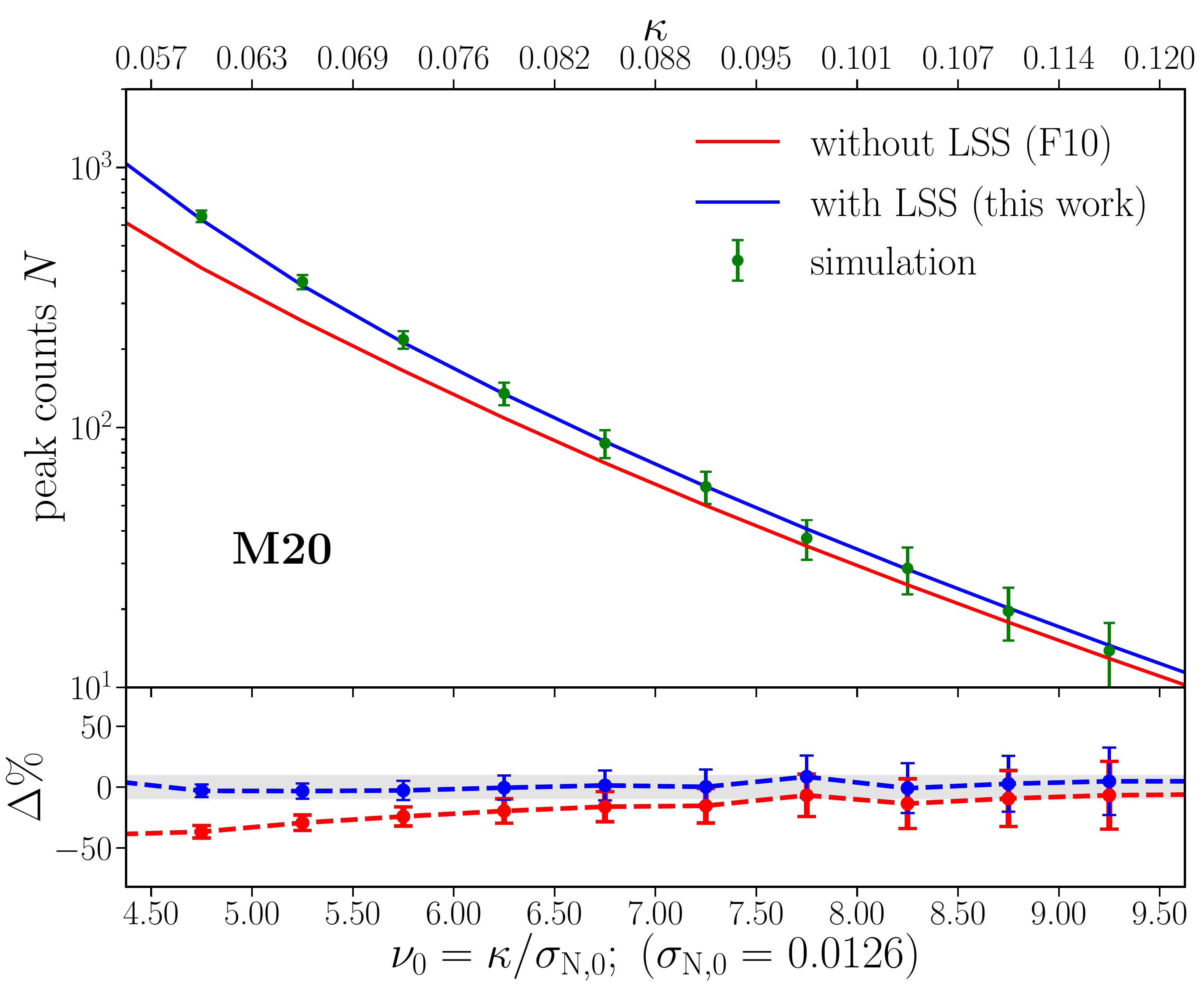}\includegraphics[width=9cm]{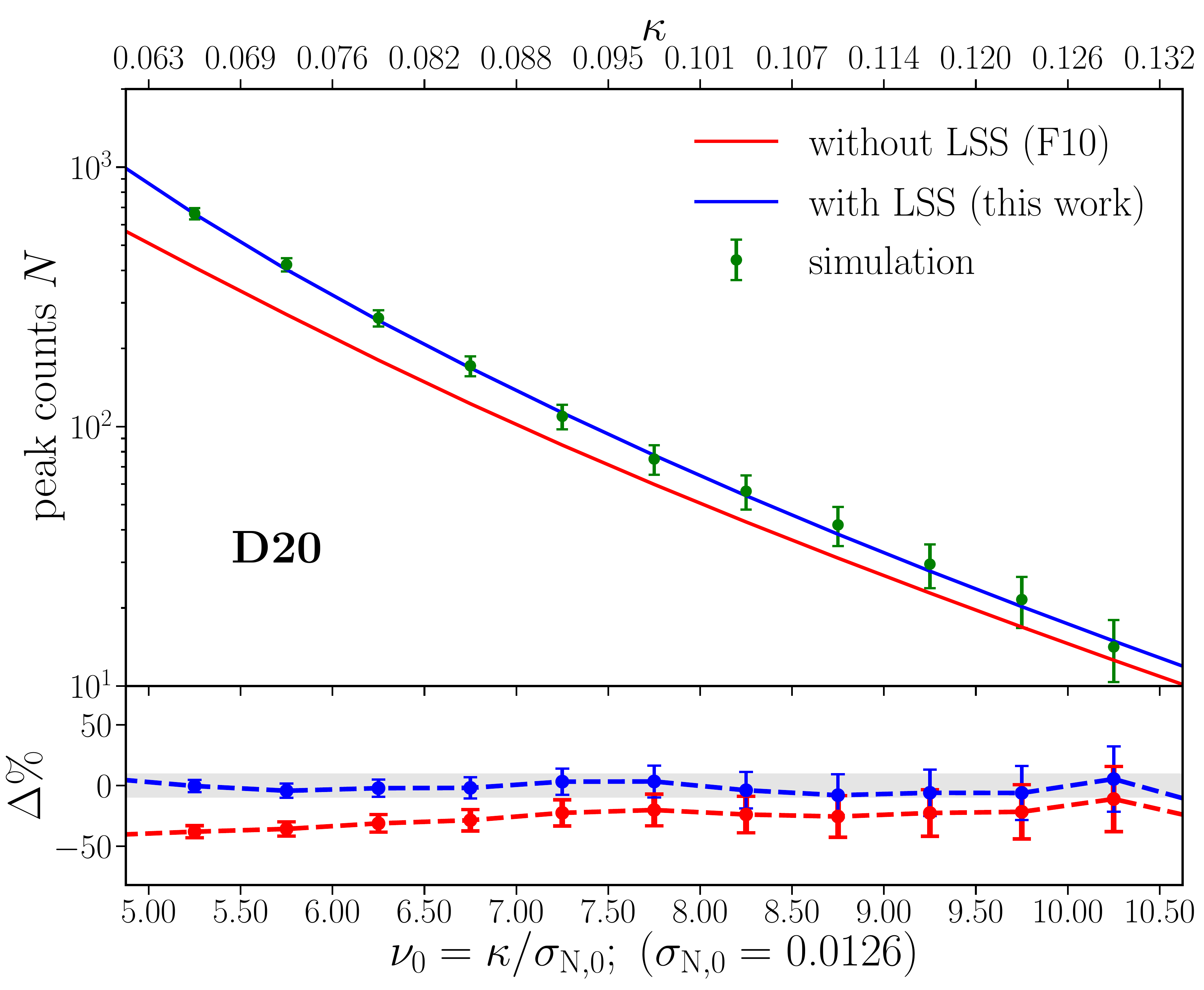} 
\par\end{centering}
\caption{\label{fig:mock_pre}
The same as in Fig.\ref{fig:mock_pre-1} but for the cases of
\texttt{ }\textbf{S10} (upper left), \textbf{S20} (upper
right), \textbf{M20} (buttom left) and \textbf{D20} (buttom right).}
\end{figure*}

The peak count comparison results are shown in Fig.\ref{fig:mock_pre-1}
and Fig.\ref{fig:mock_pre} corresponding to the survey conditions
listed in Table \ref{tab:Mocks} respectively. The green symbols are
the results averaged over the corresponding $20\times96$ maps and then scaled to the considered survey area. The
error bars are the corresponding Poisson errors. The blue lines are the results from our model including the
LSS effect, and the red lines are from \citetalias{Fan2010} without the LSS effect.
Again, the shown results are the peak counts vs. $\nu_{0}$ defined
by the shape noise $\sigma_\mathrm{N,0}$. The corresponding $\kappa$ are
indicated in the upper horizontal axes. In the bottom part of each
panel, we show the fractional differences of the two models with respect
to the simulation results.

%F7

Fig.\ref{fig:mock_pre-1} shows the results for \textbf{S10small}  with $z_{\mathrm{med}}=0.7$,
$n_{g}=10\,\mathrm{arcmin^{-2}}$, and $S=150\deg^{2}$, similar to
the current accomplished WL surveys. In this case, $\sigma_\mathrm{LSS,0}=0.0057$
and $\sigma_\mathrm{N,0}=$ 0.0178 for $\theta_{G}=2.0\,\mathrm{arcmin}$.
Thus the contribution from LSS is much smaller than that from the
shape noise, and its effect on WL peak counts is rather weak considering
relatively large error bars. This can be evidently seen from the lower
part of the panel. Both the blue and red lines agree with the simulation
results very well for high peaks with $\nu_{0}\geqslant4$ with the
fractional differences less than $10\%.$

In the upper left panel of Fig.\ref{fig:mock_pre}, we show the results
of \textbf{S10}. In this case, the number density and the redshift
distribution of source galaxies are the same as those in Fig.\ref{fig:mock_pre-1},
but with a larger survey area with $S=1086\deg^{2}$. Thus the statistical
errors of WL peak counts are smaller by $\sim2.7$ times than that
of \textbf{S10small}. We see again that both models work well, and
the model including the LSS effect (blue) gives better results for
the left two bins.

In the upper right panel of Fig.\ref{fig:mock_pre}, the results of
\textbf{S20} with $z_{\mathrm{med}}=0.7$, $n_{g}=20\,\mathrm{arcmin^{-2}}$,
and $S=1086\deg^{2}$ are presented. In this case, the LSS effect is
the same as that of \textbf{S10}, but the shape noise is lower with $\sigma_\mathrm{N,0}=0.0126$.
Thus the relative contribution of the LSS effect should be stronger
than the case of \textbf{S10}. We see that for $\nu_{0}>5$, although
both the blue line and the red line agree with the simulation results
within $10\%$, the red line is systematically lower, showing that
the LSS effect starts to be important. For $\nu_{0}<5$, the red line
deviates significantly from the simulation results, but our current
model including the LSS effect can give excellent predictions out
to $\nu_{0}\sim4$.

The results for \textbf{M20} with $z_{\mathrm{med}} =1.0$, $n_{g}=20\,\mathrm{arcmin^{-2}}$
and $S=1086\deg^{2}$ are shown in the lower left panel of Fig.\ref{fig:mock_pre}.
Here the LSS contribution increases to $\sigma_\mathrm{LSS,0}=0.0082$. Comparing
to the upper right panel, we see that the model prediction without
the LSS effect (red line) significantly underestimates the peak counts
over the whole considered range. Taking into account the LSS effect,
our improved model works very well to ${\cal K} \gtrsim 0.06$, corresponding
to $\nu_{0} \gtrsim 4.5$.

With even higher $z_{\mathrm{med}} $, the LSS projection effect gets larger.
The lower right panel of Fig.\ref{fig:mock_pre} shows the results
of \textbf{D20} with $z_{\mathrm{med}}=1.4$, $n_{g}=20\,\mathrm{arcmin^{-2}}$
and $S=1086\deg^{2}$. In this case, $\sigma_\mathrm{LSS,0}=0.0108$, is comparable
to that from the shape noise with $\sigma_\mathrm{N,0}=0.0126$. Without
including the LSS projection effect, the underestimate is at the level of $30\%$,
much larger than the statistical errors. While including the LSS effect, the
model predictions (blue lines) are in excellent agreement with the simulation results
to ${\cal K} \gtrsim 0.063$, or $\nu_{0}\gtrsim5$.

Note that for the three cases with $n_{g}=20\,\hbox{arcmin}^{-2}$,
the WL lensing signal from a halo increases with the increase of $z_{\mathrm{med}}$.
Thus we see a somewhat increase of the lower limit of $\nu_{0}$ above
which our high-peak halo model applies from $z_{\mathrm{med}}=0.7$ to $z_{\mathrm{med}}=1.4$.

\subsection{Cosmological constraints}

To demonstrate explicitly the LSS effect on the cosmological constraints
derived from WL high peak counts and how our improved model performs, here
we run MCMC fitting using WL mock data.

For \textbf{S10}, \textbf{S20}, \textbf{M20} and \textbf{D20} in Table
\ref{tab:Mocks}, we generate, respectively, the WL peak count mock
data by averaging over the $20\times96$ maps. For \textbf{S10small}, we scale
the peak counts obtained for \textbf{S10} to $S=150\deg^{2}$. The central
data points in different bins $\{N_{i}\}$ for different cases are
the same as those shown in Fig.\ref{fig:mock_pre-1} and in Fig.\ref{fig:mock_pre}.
Note that for the upper end of the peaks, we only use bins with $N_{i} \gtrsim10$.

We employ the $\chi^{2}$ fitting to constrain cosmological parameters
from WL mock data. The $\chi^{2}$ is defined as

\begin{equation}
\chi^{2}=\bm{\Delta}^{T} \widehat{\mathbf{C}^{-1}} \bm{\Delta},\label{eq:chi}
\end{equation}
where $\text{\ensuremath{\bm{\Delta}}}$ is $\bm{\Delta}\equiv\bm{N}-\widehat{\bm{N}}$
with $\bm{N}$ being the mock data vector consisting of WL peak counts
of different bins and $\widehat{\bm{N}}$ being the model predictions
for these bins. The quantity $\mathbf{C}$ is the covariance matrix
for peak counts between different bins. We calculate it
using the simulation at the fiducial cosmology. Specifically,
for each case in Table \ref{tab:Mocks}, we first obtain the covariance
for an area $S_{0}$ corresponding to an individual simulated convergence
map by calculating the variance of peaks $[\mathbf{C}_{0}]_{ij}$
between $i$th bin and $j$th bin from $20\times96$ maps. We then
scale $\mathbf{C}_{0}$ to the mock survey area $S$ considered in different
cases by

\begin{equation}
\mathbf{C}=\frac{S}{S_{0}}\mathbf{C}_{0}.
\end{equation}

Studies have shown that this scaling can lead to a slight underestimate
of the covariance for large $S$ \citep{Kratochvil2010}. This, however,
should not affect our conclusions regarding the bias resulting from
neglecting the LSS effect and the validity of our new model. It is
also noted that the cosmology-dependence of the covariance is not
considered here.

From $\mathbf{C}$, we can calculate its inverse $\mathbf{C}^{-1}$
and further the $\widehat{\mathbf{C}^{-1}}$:
\begin{equation}
\widehat{\mathbf{C}^{-1}}=\frac{R-N_{{\rm bin}}-2}{R-1}{\mathbf{C}}^{-1},
\end{equation}
where $R=20\times96$ and $N_{{\rm bin}}$ is the number of bins of
WL peak counts used in deriving cosmological constraints.

\begin{figure}[h]
\begin{centering}
\includegraphics[width=9cm]{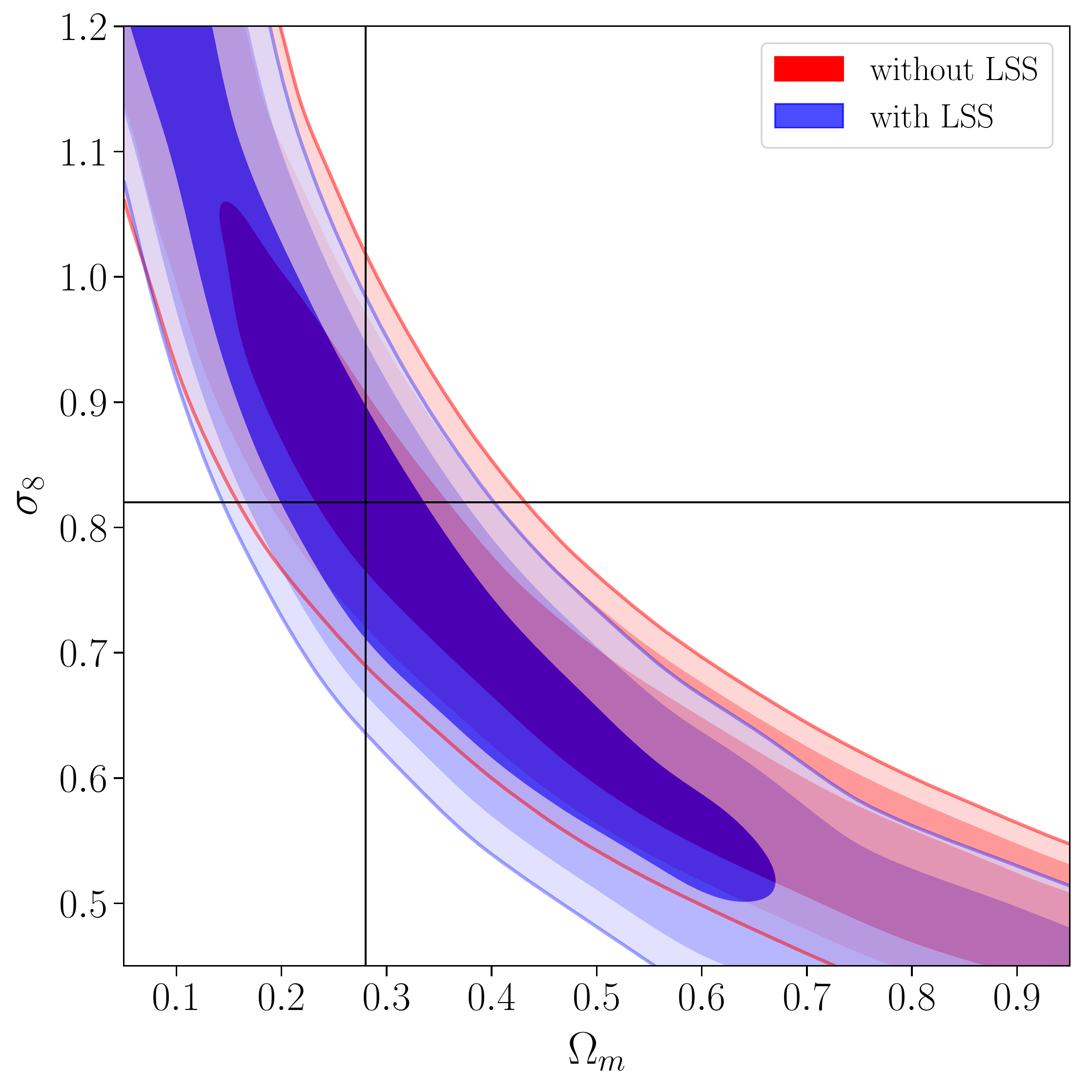} 
\par\end{centering}
\caption{\label{fig:MCMC-1}The derived cosmological constraints for \textbf{S10small}. The
red contours are the constraining results of $1-\sigma$,$2-\sigma$ and $3-\sigma$ using \citetalias{Fan2010} model and the blue ones
are from the model in this work. The input cosmological parameters for simulation is indicated by green crossing lines.}
\end{figure}

\begin{figure*}
\begin{centering} 
\includegraphics[width=8cm]{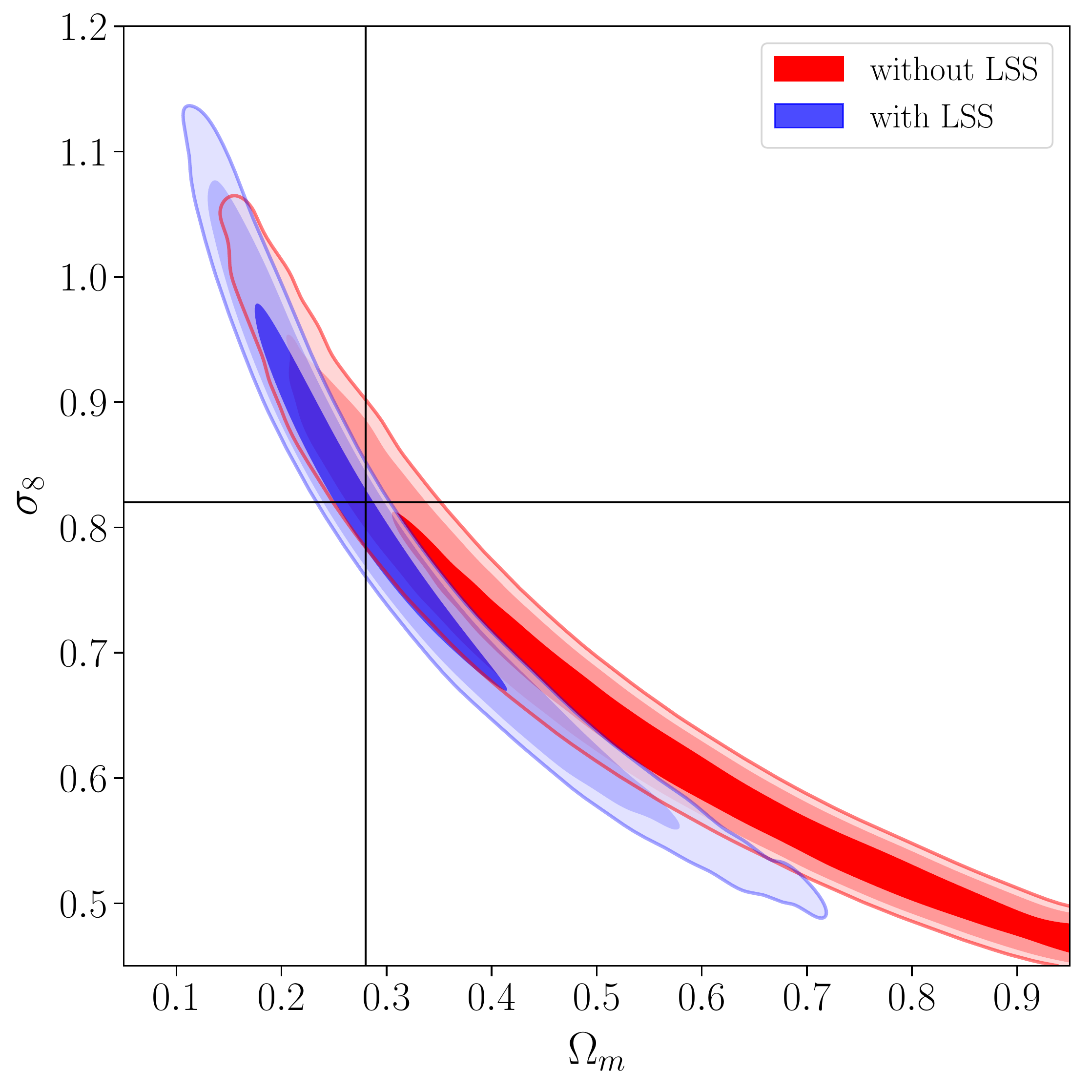}\includegraphics[width=8cm]{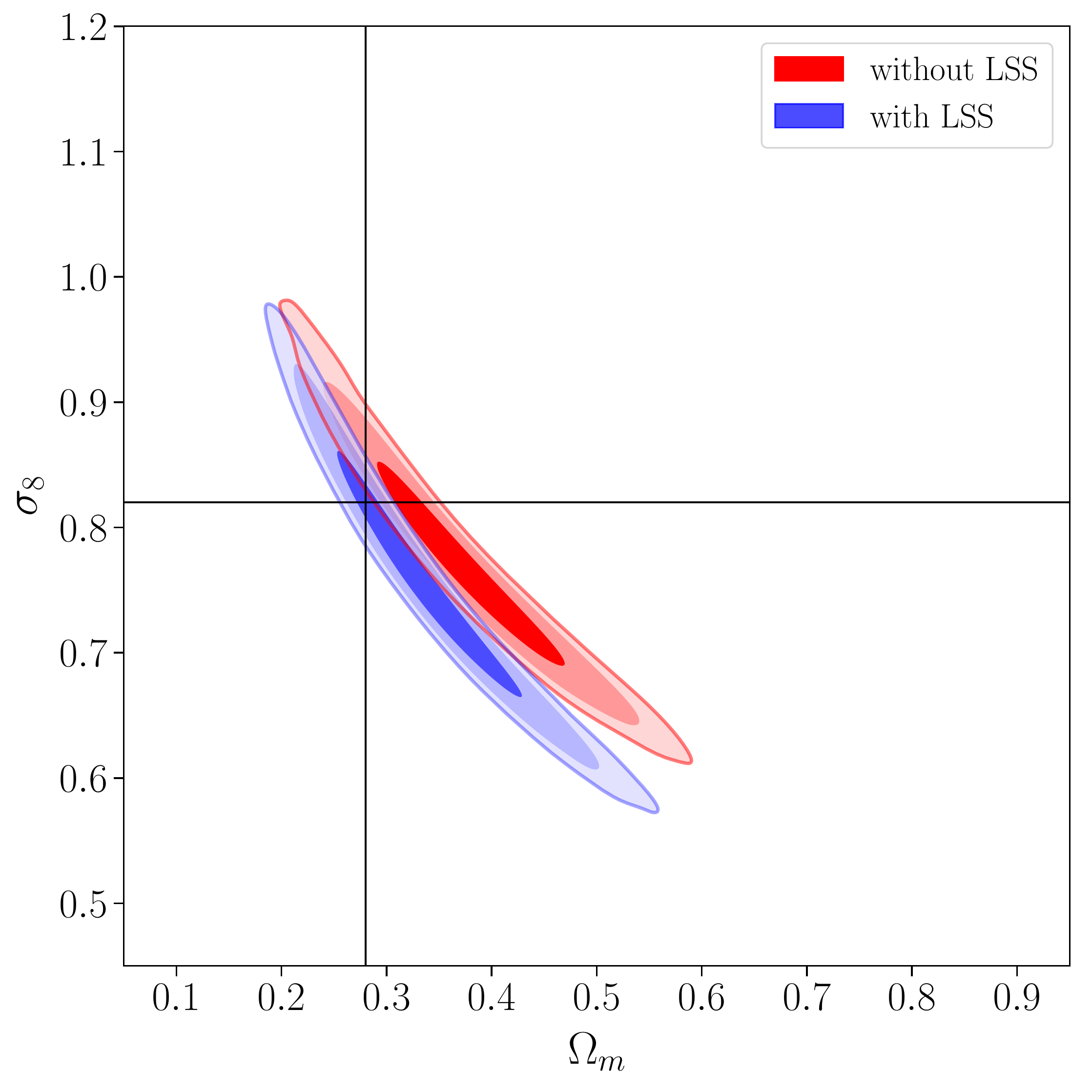} 
\par\end{centering}
\begin{centering}
\includegraphics[width=8cm]{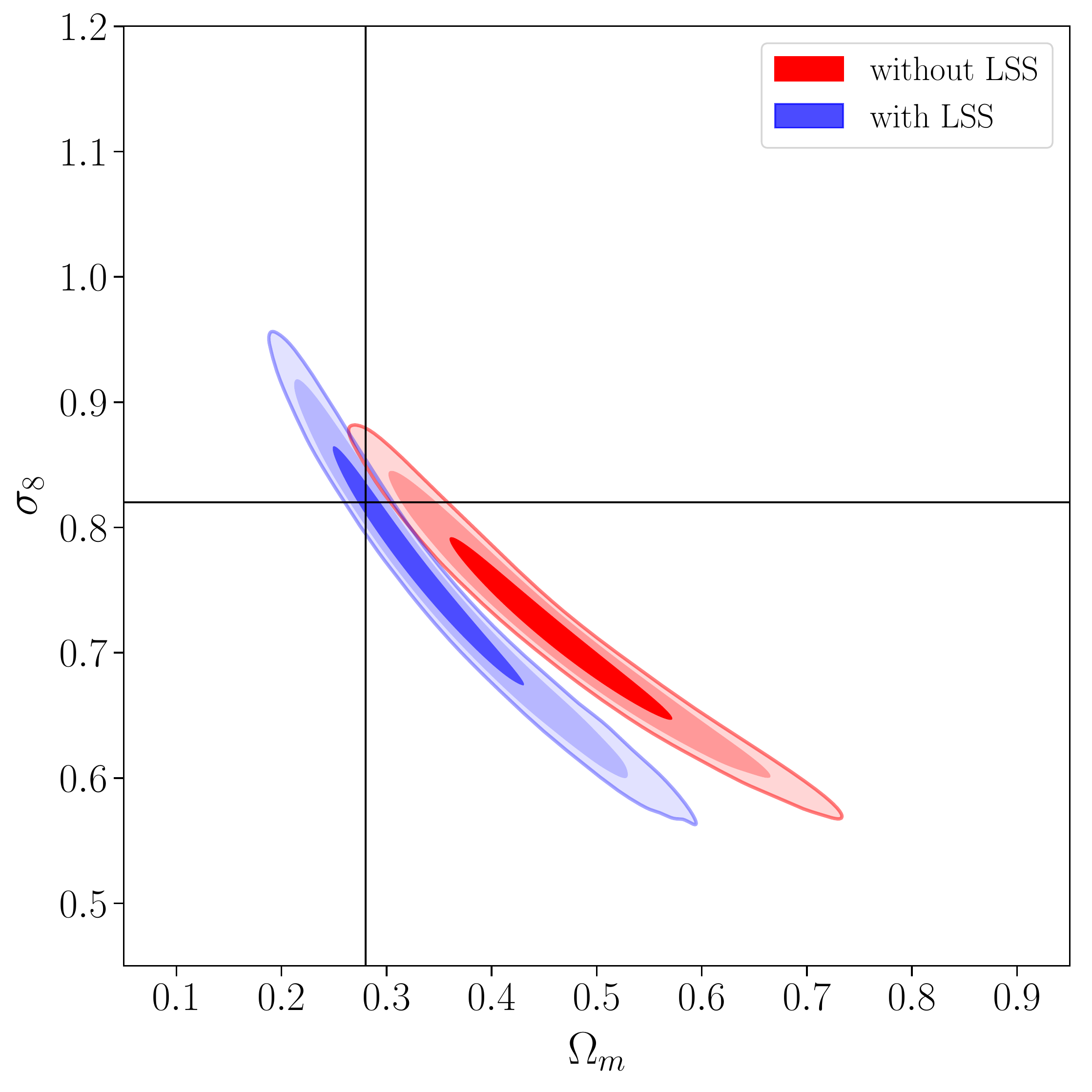}\includegraphics[width=8cm]{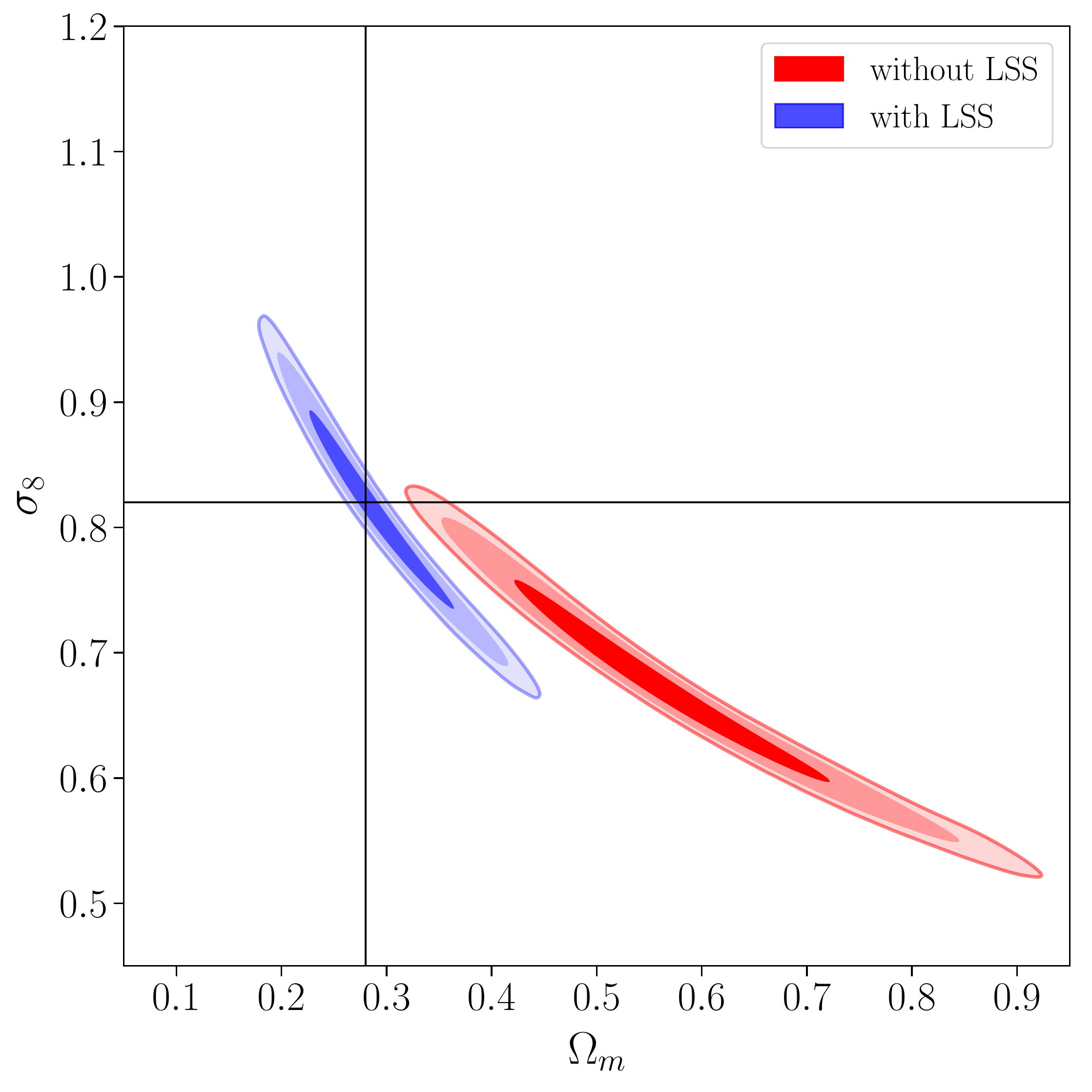} 
\par\end{centering}
\centering{}\caption{\label{fig:MCMC-4}Results for mock \textbf{S10} (upper left)\texttt{,}\textbf{S20}
(upper right)\texttt{,}\textbf{M20} (lower left) and \textbf{D20}
(lower right), respectively.}
\end{figure*}
In our analyses here, we concentrate on the constraints on $\Omega_{m}$
and $\sigma_{8}$, and set all the other cosmological parameters fixed
to be the input values of the simulations. We implement the MCMC technique
to explore the posterior probabilities of ($\Omega_{m}$, $\sigma_{8}$)
({\color{cyan}X. K. }\citealp{Liu2015,Liu2016}).

\begin{table*}
\centering{}\caption{\label{tab:dgn}Degeneracy parameter $(\alpha,\Sigma_{8})$ for different
cases. The degeneracy is defined by $\Sigma_{8}=\sigma_{8}(\Omega_{m}/0.27)^{\alpha}$.}
%\begin{tabular}{clllllll}
%\begin{tabular}{cccccccc}
\begin{tabular}{cCCCCCC}
\toprule 
Mocks  & \textbf{S10small} \footnotemark[1] & \textbf{S10small}  & \textbf{S10}  & \textbf{S20}  & \textbf{M20}  & \textbf{D20}\tabularnewline
\hline 
$\alpha$    &0.434            &0.452            &0.456            &0.493           &0.465           &0.417           \tabularnewline
%\hline 
$\Sigma_{8}$&0.833$\pm$0.045  &0.791$\pm$0.050  &0.814$\pm$0.016  &0.836$\pm$0.011 &0.838$\pm$0.009 &0.834$\pm$0.008 \tabularnewline
\hline
\end{tabular}
\footnotetext[1]{ derived from \citetalias{Fan2010} model.}
\end{table*}

%F10
\begin{figure}[h]
\begin{centering}
\includegraphics[width=9cm]{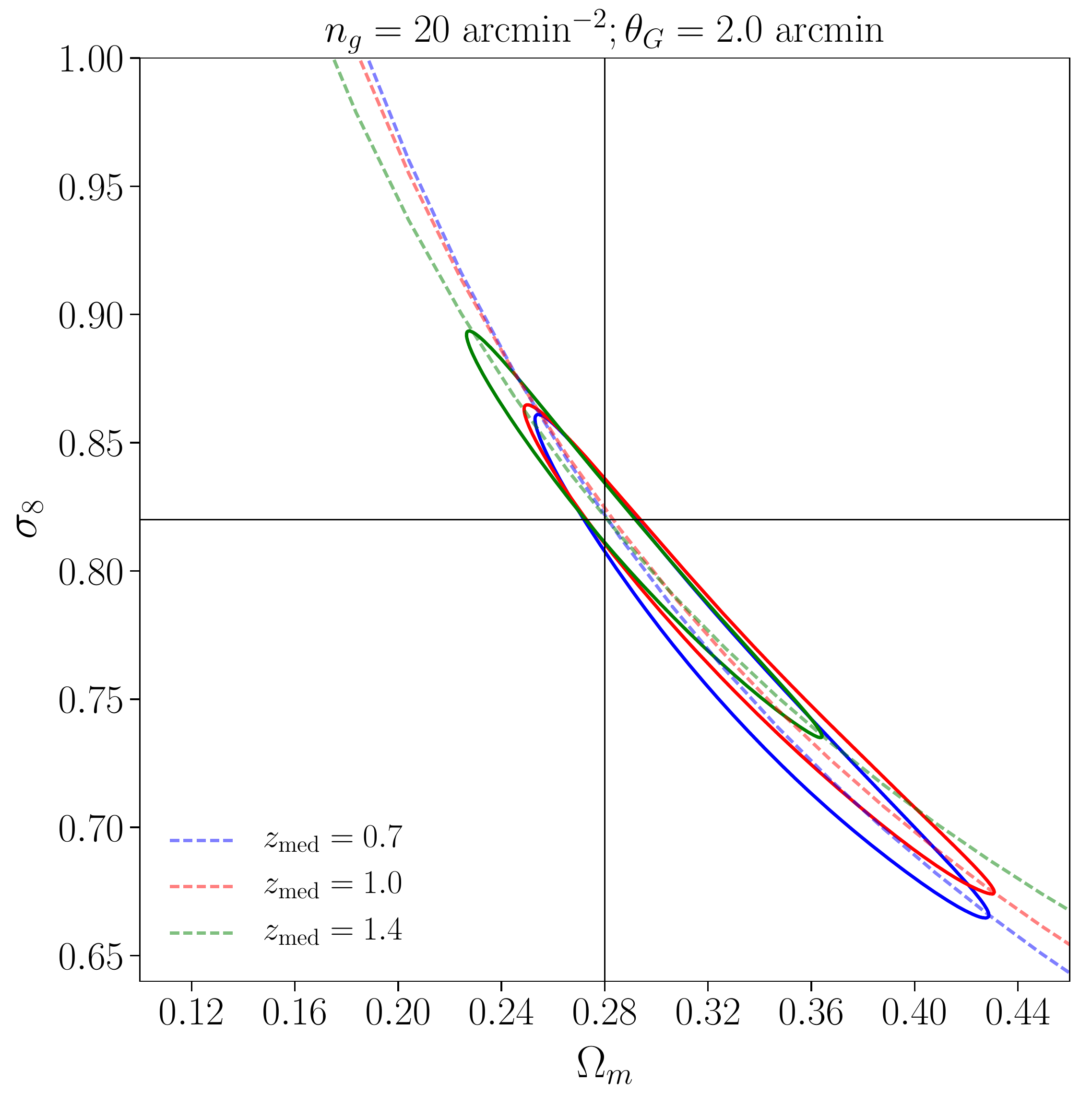}
\par\end{centering}
\caption{\label{fig:dgn}The degeneracy of $(\Omega_{m},\sigma_{8})$.
The contours are the $1\,\sigma$ regions from the peak abundance
model including LSS projection for \textbf{S20}, \textbf{M20}
and \textbf{D20}. The dashed lines
are the corresponding fitted degeneracy curves defined by $\Sigma_{8}=\sigma_{8}(\Omega_{m}/0.27)^{\alpha}$.
The input cosmological parameters for mock data are shown as the black cross.}
\end{figure}

The constraining results for \textbf{S10small} are shown in Fig.\ref{fig:MCMC-1},
where the blue and red contours are the results using the model presented
in this paper including the LSS effect and the model of \citetalias{Fan2010} without
the LSS effect, respectively. The green cross indicates the input 
values of the two parameters for WL simulations. Consistent with that shown in Fig.\ref{fig:mock_pre-1},
the two constraints overlap significantly and the two models perform
equally well. In this case, the LSS effect is negligible, and the
application of \citetalias{Fan2010} model is well justified without introducing notable
biases in the parameter constraints.

The results for \textbf{S10}, \textbf{S20}, \textbf{M20} and \textbf{D20}
are presented in Fig.\ref{fig:MCMC-4}. Because of the survey area
being larger than that of \textbf{S10small}, the statistical errors are reduced
considerably resulting smaller contours. For \textbf{S10} (upper left), the
blue and red contours still have a large overlap. The WL simulation input values
are at the edge of  the $1-\sigma$ red region. The blue constraints
from our improved model including the LSS effect, on the other hand, give better results.

For \textbf{S20} (upper right), $\sigma_\mathrm{N,0}=0.0126$, $\sigma_\mathrm{LSS,0}=0.0057$,
and the total $\sigma_{0}=0.0138$. The fractional contribution from
LSS is $\sigma_\mathrm{LSS,0}/\sigma_{0}\approx0.41$. Thus the LSS effect
is already apparent. The constraints obtained by using the model of
\citetalias{Fan2010} are biased by more than $2\sigma$. For \textbf{M20} (lower left) and \textbf{D20} 
(lower right), the shape noise is the same as that of \textbf{S20}. But the
LSS effect is stronger with $\sigma_\mathrm{LSS,0}=0.0082$ and $0.0109$,
and the corresponding fractional contribution to $\sigma_{0}$ is
$\sim0.55$ and $\sim0.65$ for \textbf{M20} and \textbf{D20}, respectively. Without
the LSS effect, the derived constraints are severely biased by more
than $3\sigma$ for \textbf{M20} and even larger for \textbf{D20}. On the other
hand, in all the cases, our new model incorporating the LSS effect
works excellently with the input values being aligned with the degeneracy direction and well within the $1\sigma$ region as shown in
blue.

It is known that WL effects depend on $\Omega_{m}$ and $\sigma_{8}$
in a degenerate way, and the derived constraints of the two parameters
are highly correlated, as seen from Fig.\ref{fig:MCMC-1} and Fig.\ref{fig:MCMC-4}.
Such a correlation is often described by a relation $\Sigma_{8}=\sigma_{8}(\Omega_{m}/0.27)^{\alpha}$.
In Table.\ref{tab:dgn}, we list the values of $\alpha$ and $\Sigma_{8}$
for different cases. 
These values are derived from the principal components analysis of the MCMC samples (for details, please see \S4.1 of \citet{2005A&A...429..383T} or \texttt{PCA} method in \texttt{getdist}\footnote{\texttt{\url{http://cosmologist.info/cosmomc}}}).
Because \citetalias{Fan2010} model works well for \textbf{S10small}, for this case, we also list the values obtained
from the constraints using \citetalias{Fan2010}. For the other cases we only show the results derived from the blue regions
in Fig.\ref{fig:MCMC-4}. We see that for \textbf{S10small}, we have $\alpha\approx0.434$
from \citetalias{Fan2010} and $\alpha\approx0.452$ from our improved model. The two
results are very similar and consistent with the one we obtained from
WL peak analyses using CS82 ({\color{cyan}X. K.} \citealp{Liu2015}). For \textbf{S20}, \textbf{M20} and
\textbf{D20}, the $\alpha$ value decreases somewhat with the
increase of $z_{\mathrm{med}}$. We show their $1-\sigma$ contours
together with the derived degeneracy directions in Fig.\ref{fig:dgn}.
This indicates the potential of tomographic WL peak analyses, for
which, we will explore in detail in our future studies. We also note
the $\alpha$ values derived from WL high peak abundances are systematically
smaller than those from cosmic shear correlations \citep{Kilbinger2015},
showing the complementary of the two types of statistical analyses.

\subsection{ {Further tests} }\label{sec:Uf}

%F11
\begin{figure*}[!ht]
\begin{centering}
\includegraphics[scale=0.5]{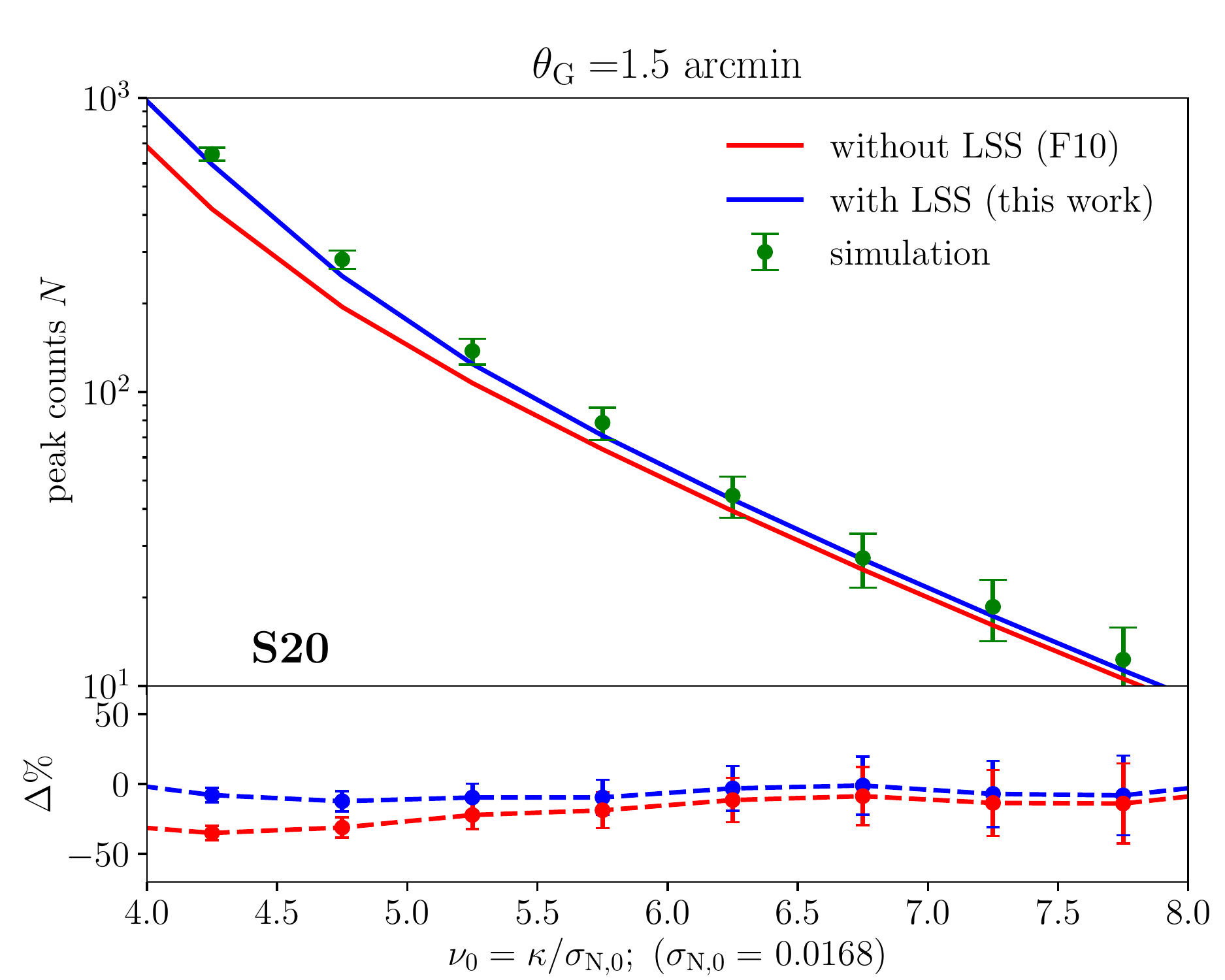}\includegraphics[scale=0.5]{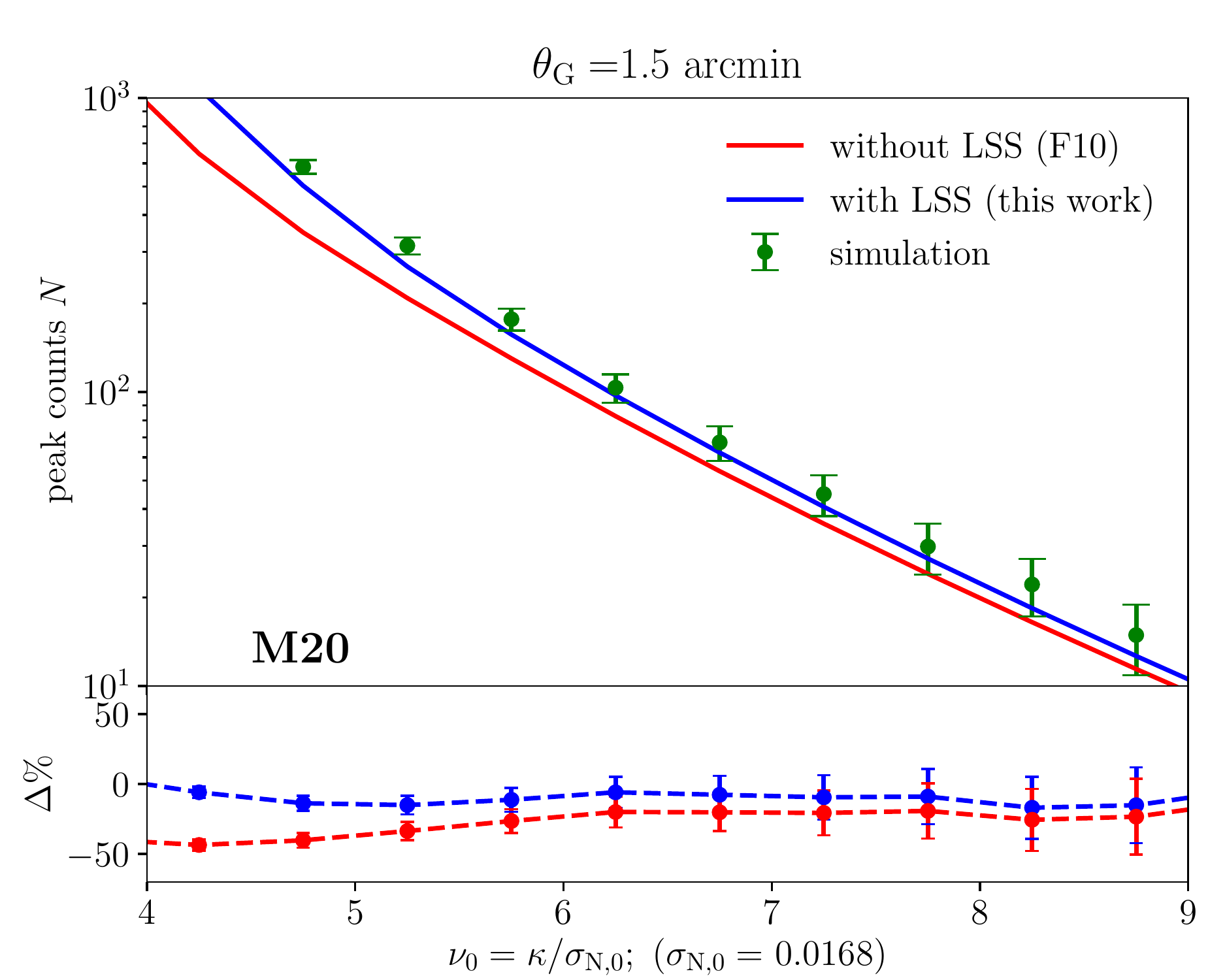}
\par\end{centering}
\begin{centering}
\includegraphics[scale=0.5]{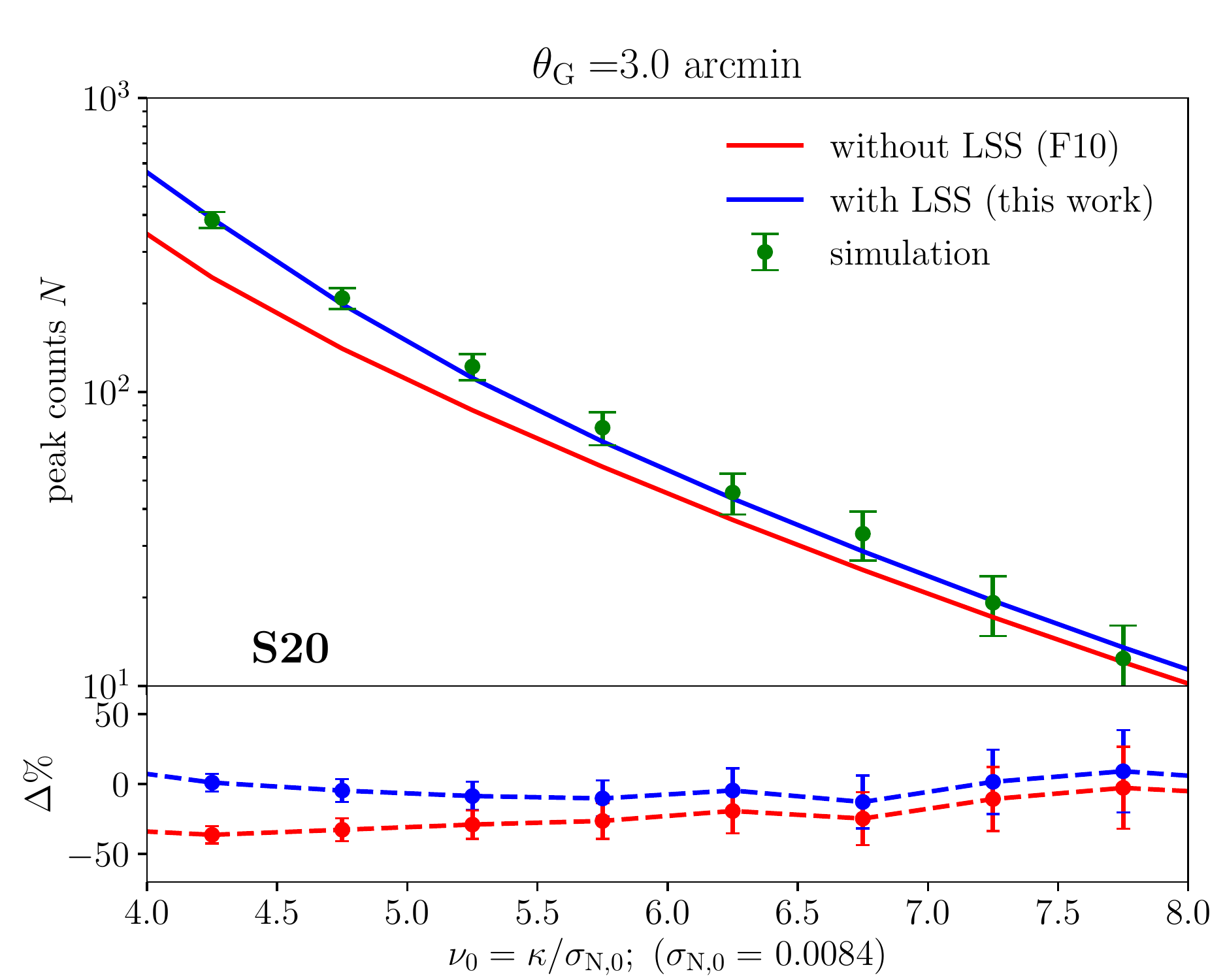}\includegraphics[scale=0.5]{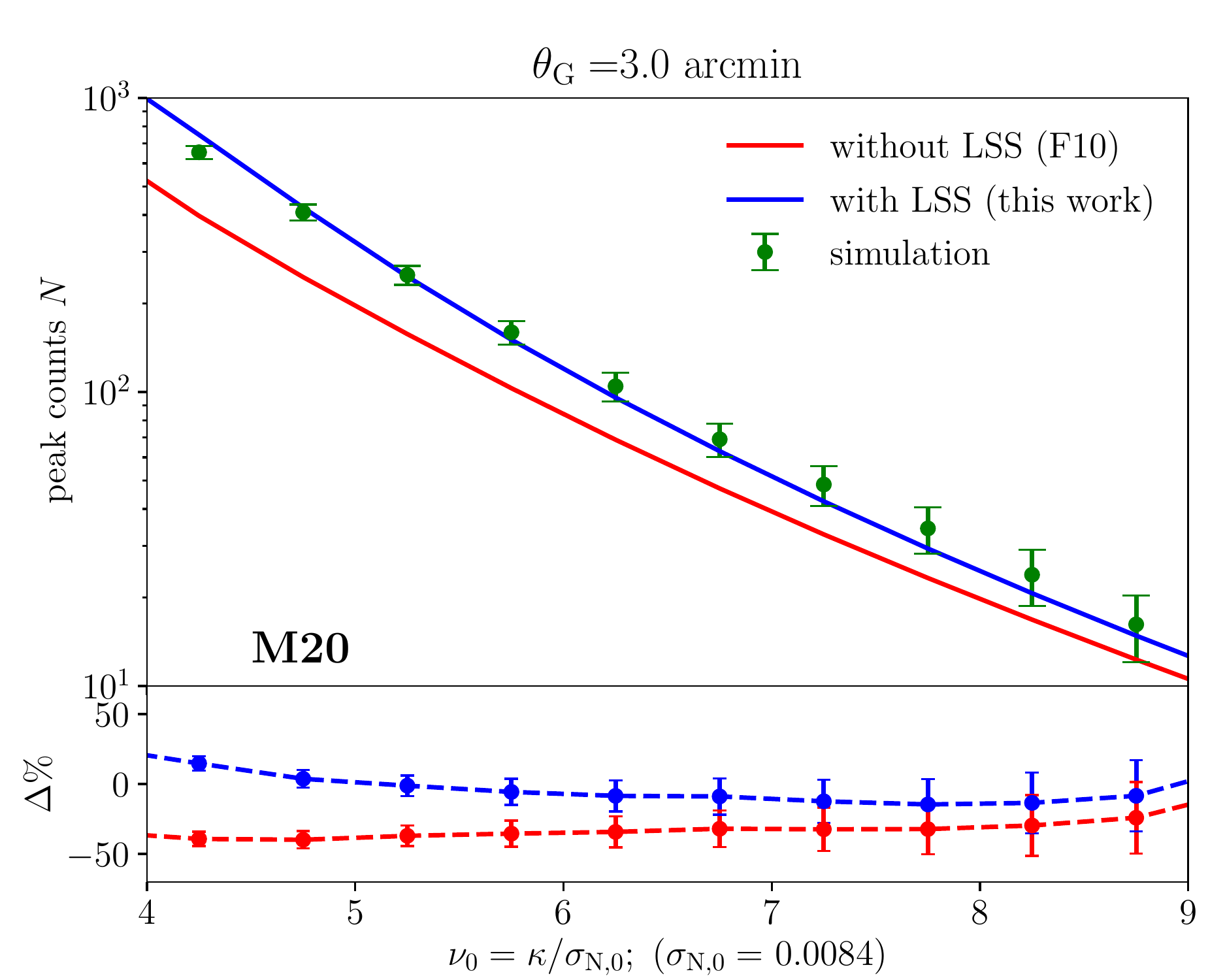}
\par\end{centering}
\caption{\label{fig:G-theta}
{
Gaussian smoothing with $\theta_{G}=1.5\, \mathrm{ arcmin} $ (upper) and $\theta_{G}=3.0\, \mathrm{arcmin} $ (lower) for S20 (left) and M20 (right).
}
}
\end{figure*}

{The previous analyses show the results with a set of fiducial parameters. In this subsection, we test the validity of the model for different cases. In Fig.\ref{fig:G-theta}, we show the results with Gaussian smoothing of different smoothing scales with $\theta_G=1.5 \hbox{ arcmin}$ (upper), and $3 \hbox{ arcmin}$ (lower), respectively for S20 in left and M20 in right. We see the model performs equally well as that of $\theta_G=2 \hbox{ arcmin}$. 
}

\begin{figure}
\begin{centering}
\includegraphics[width=9cm]{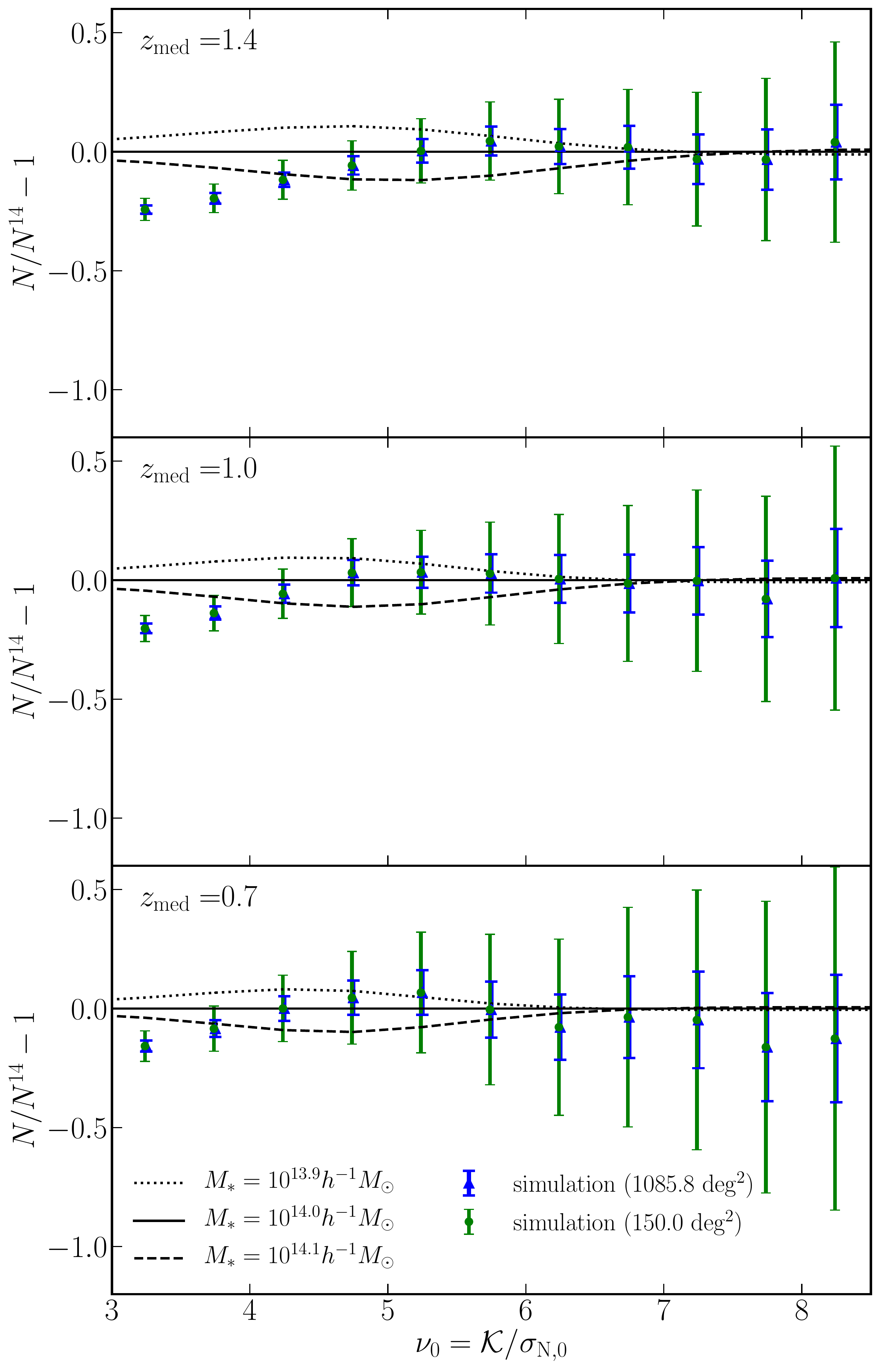} 
\par\end{centering}
\caption{
\label{fig:mstar}
{
The differences of the model predictions with $M_*=10^{13.9} h^{-1}  M_{\odot}$ and $M_*=10^{14.1} h^{-1}  M_{\odot}$ with respect to the fiducial model. The data points are the differences of the simulation results with respect to the fiducial model. The large and small error bars correspond to the survey area of $150\,\mathrm{deg}^2$ and $1086\,\mathrm{deg}^2$, respectively.
}
}
\end{figure}

{In our model, we consider massive halos with $M\ge M_*$ as the 
dominant sources of high peaks. 
Simulations show that $M_*\sim 10^{14}  h^{-1}\hbox{M}_{\odot}$ is an 
appropriate choice. 
The very precise value can have dependences on, e.g., the halo mass function 
and cosmological models. In our fiducial analyses, we take
$M_*=10^{14} h^{-1}\hbox{M}_{\odot}$. 
To test the $M_*$ sensitivity of our model predictions, in Fig.
\ref{fig:mstar},  we show the differences of the model predictions with
 $M_*=10^{13.9} h^{-1}\hbox{M}_{\odot}$ and $10^{14.1} h^{-1}\hbox{M}_{\odot}
 $ with respect to that of the fiducial results. 
The data points are the differences between the simulation results and the fiducial model predictions with  $M_*=10^{14} h^{-1}\hbox{M}_{\odot}$. The large and small error bars are for the survey area of $\sim 150 \hbox{ deg}^2$ and $\sim 1086 \hbox{ deg}^2$, respectively.
 It is seen expectedly 
 that different choice of $M_*$ has no impact on the predicted abundance of 
 very high peaks. 
 For peaks around $\nu_0 \approx 4$ in the considered cases, they show some 
 effects. 
For surveys of $\sim 150 \hbox{ deg}^2$  and $z_{\rm{med}}=0.7$, the differences arising 
 from a 0.1dex variation of $M_*$ are within the statistical errors 
 for $\nu_0 \ge 4$. 
 For surveys of $\sim 1000\, \hbox{deg}^2$, 
 or higher $z_{\rm med}$,
 the dependence on $M_*$ becomes 
 significant. We will investigate in more details on this issue in our future 
 studies.
 }

{In Fig.\ref{fig:mstar}, we extend the horizontal axis to $\nu_0=3$. 
We see that 
at $\nu_0 \sim 3$, there are some  deviations between the model predictions  and the simulation results, the higher the $z_\mathrm{med}$, the larger the deviations.
 From Fig.\ref{fig:oldvsnew}, we see that in the considered cases,
for peaks of $\nu_0 \sim 3$, a significant fraction of them are from the 
field regions resulting from the combined effects of LSS and shape noise.
Thus they are more sensitive to the LSS properties than high peaks that are 
mainly from halo regions. The Gaussian approximation of the LSS effects needs 
to be improved to better account for these relatively low peaks, particularly 
for higher $z_{\rm med}$ where the LSS effects are comparable or even larger 
than the shape noise effects. This is another important effort in our future 
studies.
}

{It is noted that our analyses here are done with the convergence fields from simulations directly. On the other hand, observations measure the shape 
ellipticities of galaxies, which directly give rise to an estimate of the reduced shear $g_i=\gamma_i/(1-\kappa)$. To perform peak analyses in the convergence 
fields, in general, we need to first reconstruct them from the shear estimates using the relation between $\gamma$ and $\kappa$. To avoid the reconstructions 
that may introduce systematic errors, the aperture mass $M_{\rm ap}$ statistics has been proposed with
(e.g., \citealp{1996MNRAS.283..837S,2004MNRAS.352..338J,1998A&A...334....1V}),
}

{\begin{equation}
M_{\mathrm{ap}}(\vec {\theta})  = \int \mathrm{d}^2{\vec {\theta'}} Q ( |\vec \theta-\vec {\theta'}|) g_t(\vec {\theta'}),
\end{equation}}
{where $g_t$ is the tangential component of $g$ with respect to $\vec{\theta} -\vec{\theta'}$. In the regime of $\kappa\ll1$ and $\vec g\approx \vec \gamma$, $M_{\rm ap}$ is equal to applying a $U$ filter to the $\kappa$ field with }

{\begin{equation}
Q({\theta})=-U({\theta})+\frac{2}{{\theta}^2}\int \theta' \mathrm{d} \theta' U(\theta').
\label{UQ}
\end{equation}
It is required that the $U$ filter is compensated with $\int d^2\vec \theta U(\vec \theta)=0$. Here we present the peak analyses results for $M_{\rm ap}$ obtained by applying an $U$ filter to the simulated convergence fields to show the applicability of our model. We choose a particular filter set with \citep{1998A&A...334....1V,2004MNRAS.352..338J}:
}
\begin{equation}
U(\theta,\theta_{U})=\frac{1}{\pi\theta_{U}^{2}}\left(1-\frac{\theta^{2}}{\theta_{U}^{2}}\right)\exp\left(-\frac{\theta^{2}}{\theta_{U}^{2}}\right).
\label{UQ}
\end{equation}

{
This filter has smoothed behaviours both in real and in Fourier spaces, and can be handled computationally better than sharply truncated filters \citep{1998A&A...334....1V}.
}

\begin{figure}
\begin{centering}
\includegraphics[scale=0.6]{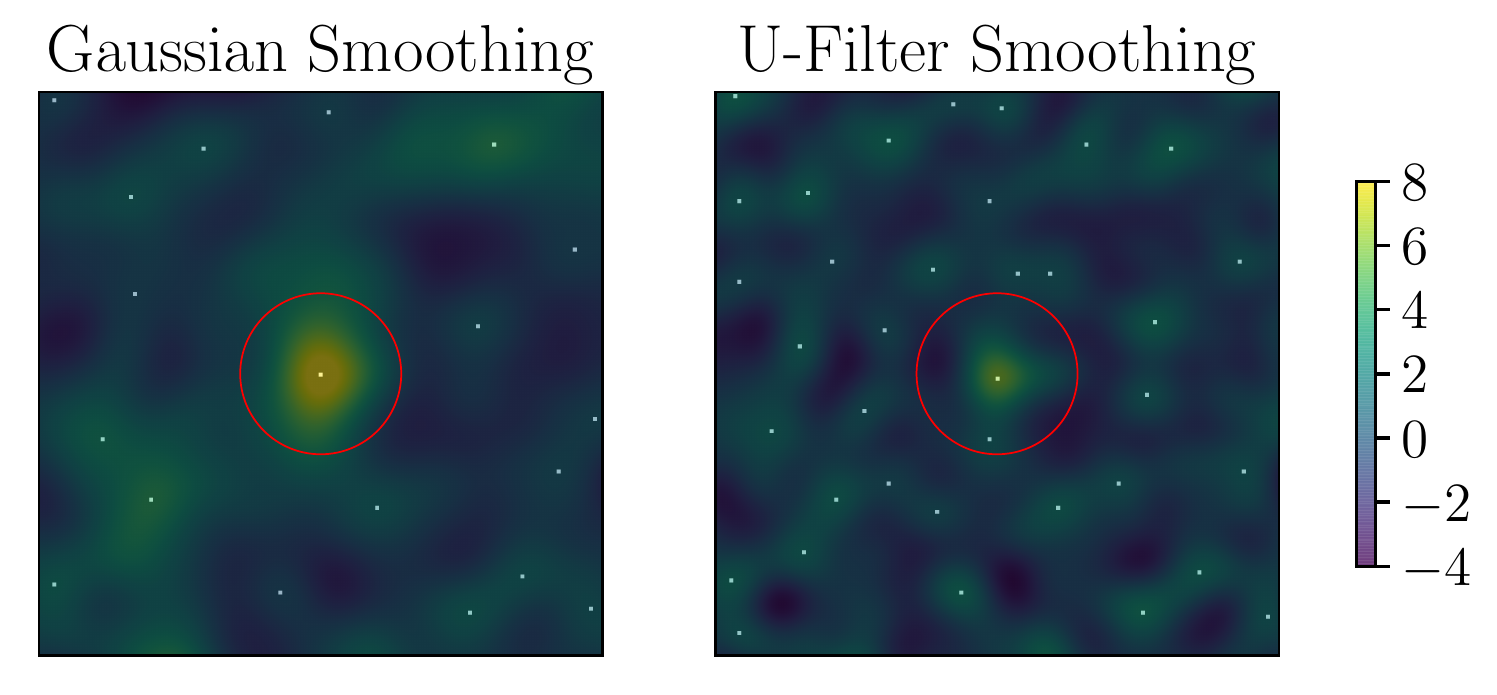}
\par\end{centering}
\caption{\label{fig:GUmatchmap} 
{A zoomed-in coomparison of convergence maps under Guassian smoothing (left) and under U filter smoothing (right) of a same field. 
The smoothing scale is $\theta_G=\theta_U=2\,\mathrm{arcmin}$. The color bar for $\nu_0$ in the two cases is shown on the right.}  
}
\end{figure}

\begin{figure}
\begin{centering}
\includegraphics[scale=0.45]{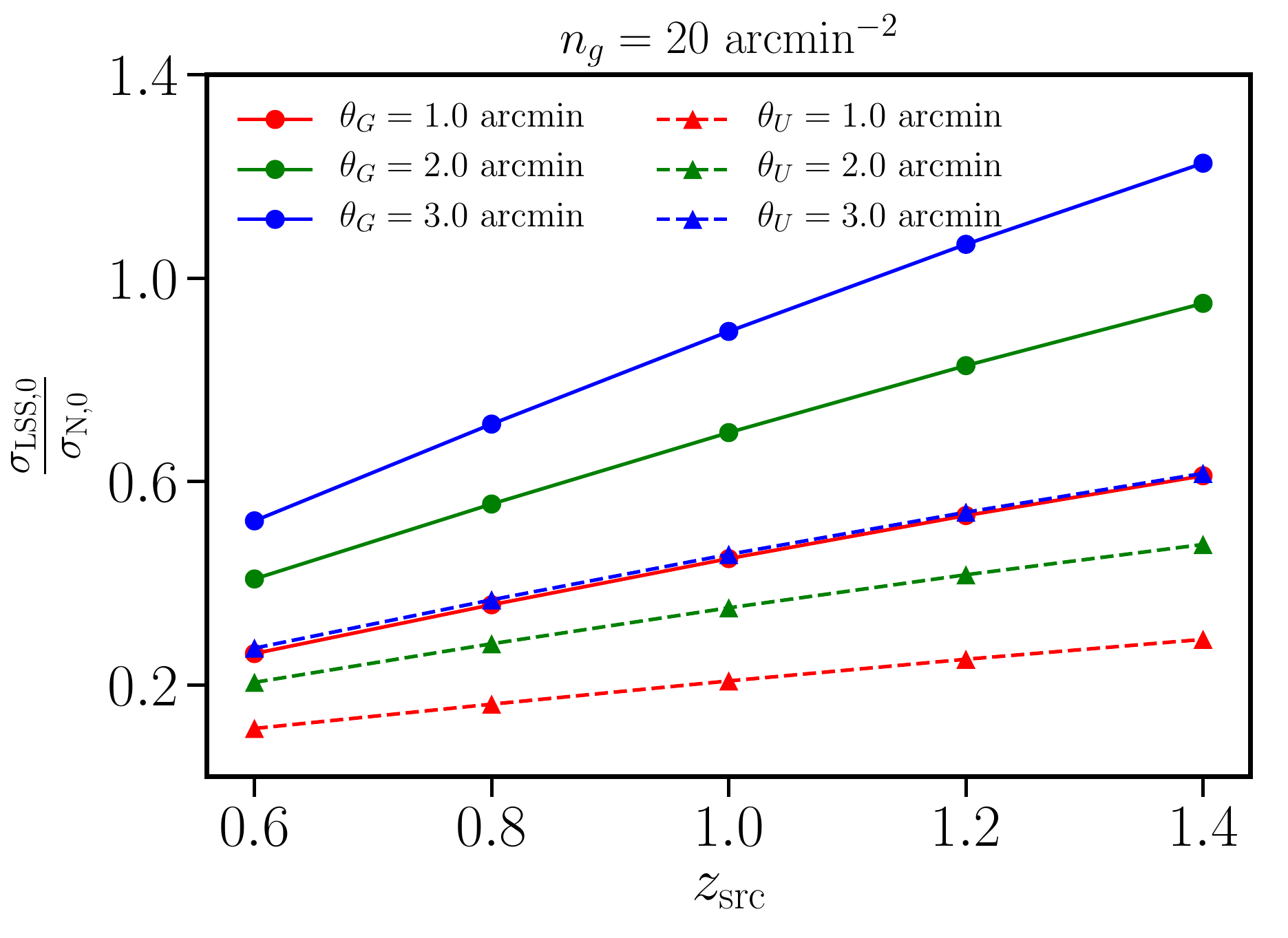}
\par\end{centering}
\caption{\label{fig:lss2n}
{
The ratio of $\sigma_{\mathrm{LSS,0}}/\sigma_{\rm{N,0}}$ for Gaussian (solid) and U (dashed) filterings at different source redshifts and different smoothing scales.
}
}
\end{figure}

\begin{figure}
\begin{centering}
\includegraphics[scale=0.8]{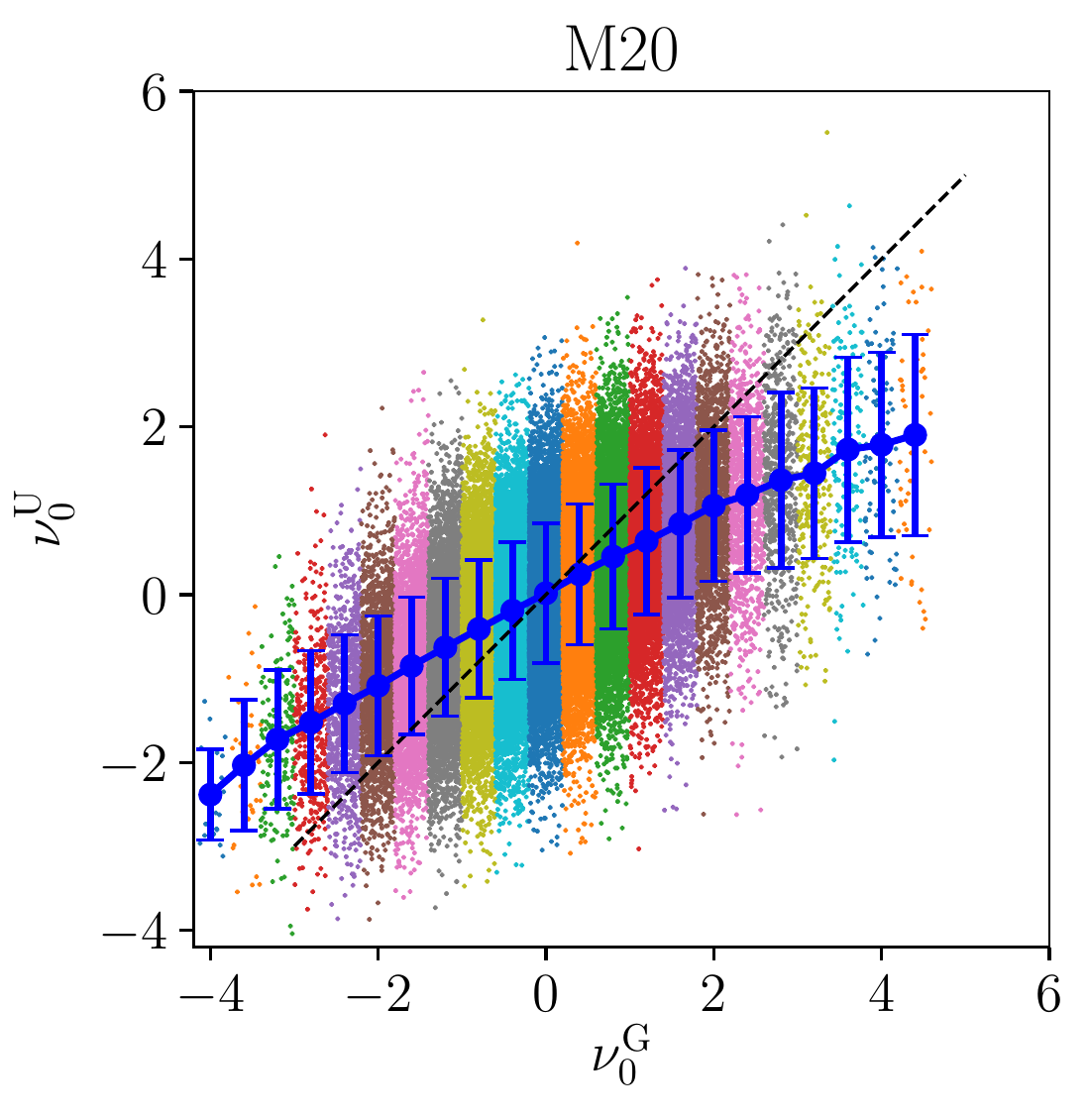}
\par\end{centering}
\caption{\label{fig:GUmatchSNR}
{The scatter plot of $\nu_0^{\mathrm{G}}-\nu_0^{\mathrm{U}}$ relation for $\theta_{G}=\theta_{U}=2.0$ arcmin applied to M20 mocks.} 
}
\end{figure}

\begin{figure}
\includegraphics[scale=0.48]{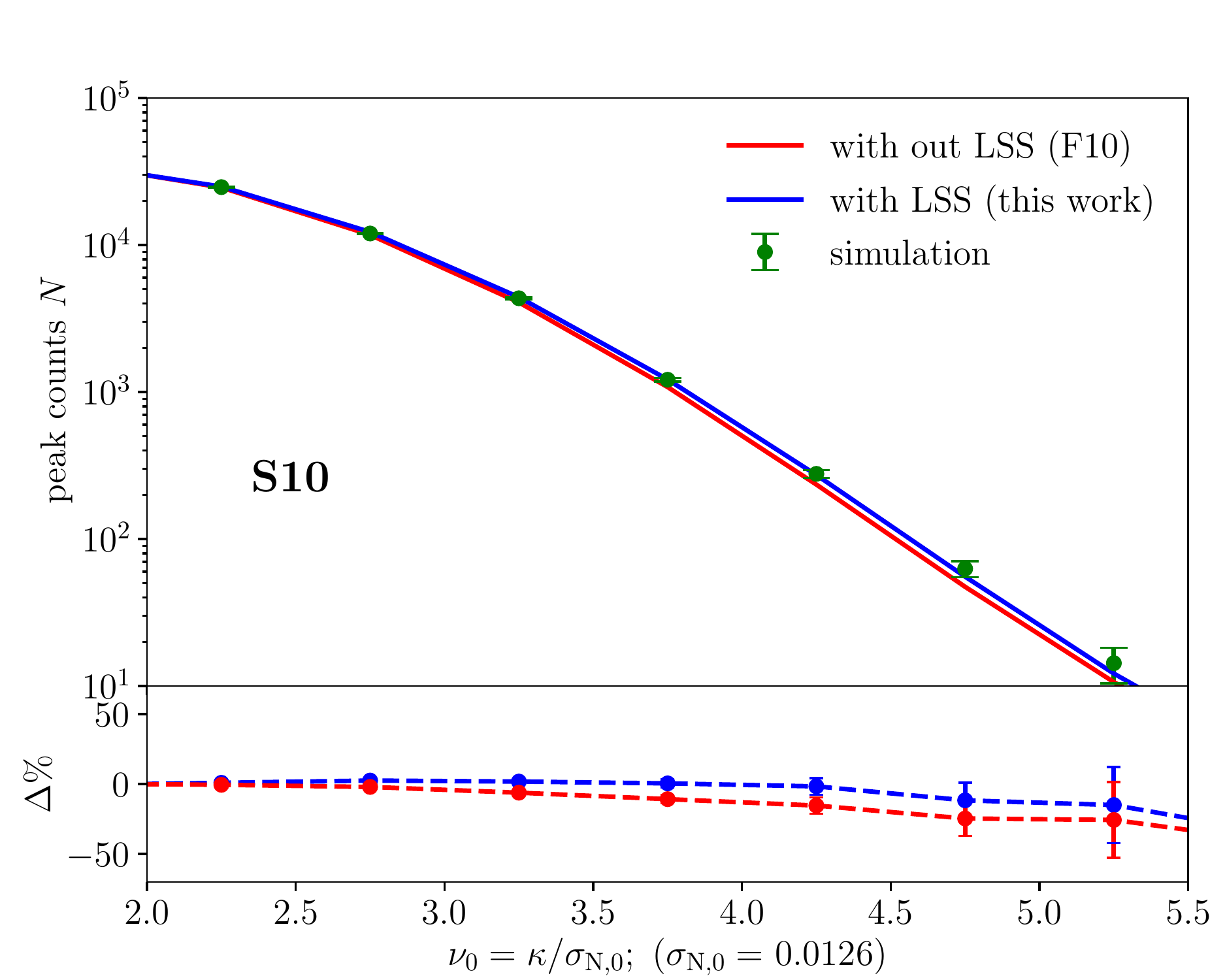}
\includegraphics[scale=0.48]{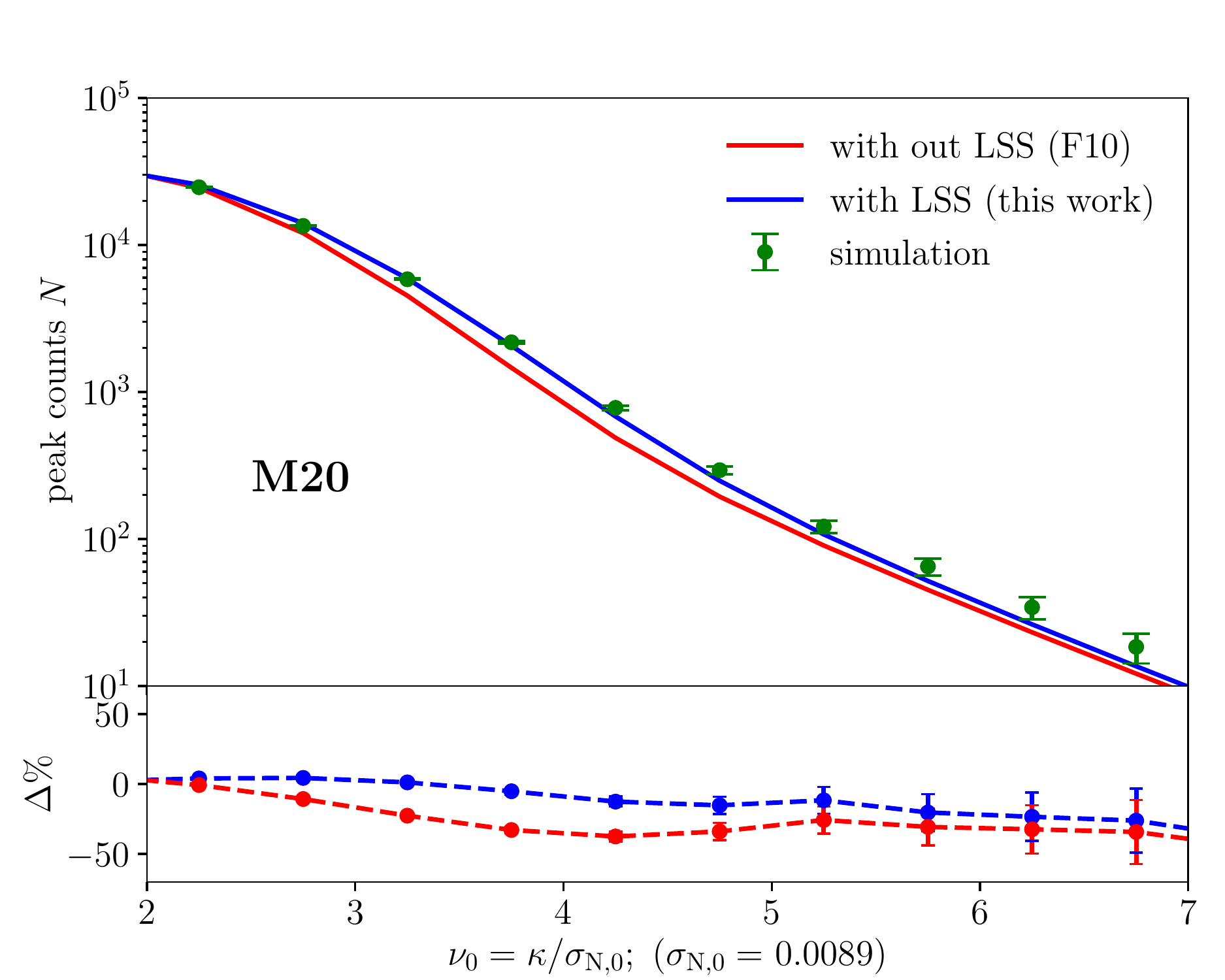}
\caption{\label{fig:npk-u}
{Aperture peak abundance for S10 and M20. The U-filter smoothing scale is $\theta_\mathrm{U}=2.0\,\mathrm{arcmin}.$
}}
\end{figure}

{
In Fig.\ref{fig:GUcl}, we already show the comparison of  power spectra in Gaussian and in U filters. The U filter can filter out the large-scale contributions more efficiently than that of the Gaussian smoothing. For a visual comparison, we show in Fig.\ref{fig:GUmatchmap} the zoom-in maps of the Gaussian and the U filters of a same field. We see that high peaks correspond well in the two cases. On the other hand, large-scale patterns are more apparent in the Gaussian-smoothed map.  
In Fig.\ref{fig:lss2n}, we show $\sigma_{{\rm LSS},0}/\sigma_{{\rm N},0}$ for the two filters for different source redshifts and different smoothing scales. While the LSS effect increases with the source redshift and the smoothing scale in both cases, it is more significant in the Gaussian-filter case than that of the U filter, consistent with the analyses shown in  Fig.\ref{fig:GUcl} and Fig.\ref{fig:GUmatchmap}.
}
{In Fig.\ref{fig:GUmatchSNR}, we show the signal-to-noise ratio comparison of the corresponding peaks under the two filters.
We see that in general, the signal-to-noise ratio is lower in $U$ filter  than that in the Gaussian smoothing, which indicates that our peak model can be applicable to lower peaks in the $U$-filter case. Fig.\ref{fig:npk-u} presents the peak number distribution of $M_{\rm ap}$ under the $U$ filter from simulations and also from our model prediction. The results demonstrate that our model works well too in this case, starting from $\nu_0  \approx 2$. 
}

 %F12

%F13

%F14

%F15

%F16

{We should note that the equivalence of obtaining $M_{\rm ap}$ from $U$-filtering of the convergence fields with its original definition is approximate under the assumption of $g\sim \gamma$. For high peaks, however, such an approximation is not accurate enough. 
Thus to model high aperture-mass peaks better corresponding to real observational analyses,
 we need to work on the true $M_{\rm ap}$ fields derived directly by applying $Q$ filter to the reduced shear field $g_t$, which is much more computationally complicated and intensive. Great efforts have been devoted to build such a model and we will present it in our forthcoming paper by  \cite{pan18}.
}

\section{Discussion}

In this paper, we analyze the projection effect from stochastic LSS on WL high peak abundances.
Similar to \citetalias{Fan2010}, we assume that high peaks are dominantly from individually
massive halos with $M\geqslant M_{*}$. To improve \citetalias{Fan2010}, we include
the LSS effect as a Gaussian random field, and its power spectrum
$C_{\ell}^{\mathrm{LSS}}$ is calculated by subtracting the one-halo contribution
from halos with $M\geqslant M_{*}$ from the overall  non-linear power
spectrum. In other words, in our modeling, we treat the heavily non-Gaussian
contributions from massive halos to WL peaks separately using their
halo mass function and the density profiles. The rest of the line-of-sight
projection effect is regarded as the LSS effect modeled as a Gaussian
random field. We comment that in line with the halo model, $C_{\ell}^{\mathrm{LSS}}$
contains contributions from one-halo terms of smaller halos with $M<M_{*}$
and the two-halo terms between all the halos. It is also noted the
exclusion of the one-halo terms from halos with $M\geqslant M_{*}$ is important
in calculating $C_{\ell}^{\mathrm{LSS}}$ correctly. Otherwise, the LSS projection
effect would be overestimated.

To exam our model performance, we carry out extensive simulation studies
by generating WL maps with respect to different survey conditions.
Our analyses show that for a CFHTLenS-like surveys (\textbf{S10small}),
the LSS effect on WL high peak counts and subsequently the derived
cosmological constraints is negligible. This is due to its relatively
small contribution to $\sigma_{0}$ in comparison to that of the shape
noise and the large statistical errors resulting from a small sky
coverage. With the same $n_{g}=10\,\hbox{arcmin}^{-2}$ and $z_{\mathrm{med}}=0.7$
but increasing the survey area to $\sim1086\deg^{2}$ (\textbf{S10}), the \citetalias{Fan2010} model gives rise to constraints that are at 
the edge of $1\sigma$ contour. Keeping the same survey depth with $z_{\mathrm{med}}=0.7$
but increasing $n_{g}$ to $20\,\hbox{arcmin}^{-2}$, the LSS projection effect
becomes notable. Further increasing the survey depth represented by
increasing $z_{\mathrm{med}}$, the LSS projection effect gets more and more important.
With the \citetalias{Fan2010} model without the LSS projection effect, the cosmological constraints
derived from WL high peak counts are biased by more than 
$2\,\sigma$ and $3\,\sigma$ for \textbf{S20} and \textbf{M20}, and even larger for \textbf{D20}. This shows clearly that for future large WL surveys,
the LSS projection effect on WL high peaks must be taken into account. Our model
presented in this paper performs very well in catching up the effect.
We address that in our improved model, $C_{\ell}^{\mathrm{LSS}}$ contributes
additional cosmological information.

{We also show the good performance of the model for different smoothing scale $\theta_{\rm G}$ and for the aperture-mass peaks with a compensated $U$ filter. For the latter, we should keep in mind that the true aperture-mass fields are calculated from the reduced shears $g$ rather than from the shears $\gamma$.  }

We note that the mass function and the density profile of halos are
important ingredients in our model calculations. Their uncertainties
can potentially affect the model predictions. In the analyses here,
we take the halo mass function from \citet{Watson2013} and the
NFW halo density profile with the mass-concentration relation of Eq.(\ref{eq:cm}).
They work well in our comparisons with simulated high peak counts.
For future very high precision studies, we need to consider these
uncertainties more carefully. Considering the complicated mass distributions in real halos,  there should be 
a negtive bias ($\approx 10\%$) in the 2D weak-lensing-derived $M$-$c$ relation
with respect to that of 3D \citep{2015ApJ...814..120D}.
As a test, we reduce the $A$ value
in the mass-concentration relation by $10\%$, the theoretical predictions
for high peak counts decrease at the level of $\sim10\%$ for the
highest bins in Fig.\ref{fig:mock_pre}, and smaller for lower bins.
This is still within the statistical uncertainties of the peak counts
in our considered cases here with $\sim1100\deg^{2}$. With $S$ being
$\sim15,000\deg^{2}$ for future surveys, such as LSST \citep{LSST2012}
and Euclid \citep{Amendola2013}, highly accurate knowledge about
these ingredients is needed for precision studies. On the other hand,
self-calibrated approaches are possible to constrain, e.g, the mass-concentration
relation, simultaneously with cosmological parameters from WL peak
counts ({\color{cyan}{X. K. }}\citealp{Liu2015}). We will investigate these issues in detail
in our future studies.

It is also noted that our model applies to high peaks for which the
signals are mainly from single massive halos. On the other hand, simulations
have shown that low peaks also contain important cosmological information.
It is highly desirable to build theoretical models for them. For low
peaks, such as that shown in Fig.\ref{fig:LOSsmallpk}, however, we
cannot find a single halo that contributes dominantly to the peak
signal. Thus as one of our important future tasks, we need to explore
different approaches to model the low/medium peaks.

WL peak analysis has shown its power in cosmological studies. Ongoing
and future WL surveys will increase the data in quantity by orders
of magnitude comparing to that we currently have. This will lead to
a tremendous increase of the statistical power of WL studies. Meanwhile,
however, much tighter systematic error controls are needed. Besides
the LSS projection effect on WL peaks studied in this paper, there are other
systematics that we need to understand carefully, such as the intrinsic
alignments of source galaxies, photometric redshift errors, baryonic
effects, etc.. Fully exploring the complementarity of WL peak analyses
and cosmic shear correlations, not only on cosmological constraints,
but also on different responses to systematics, is also an important
and exciting direction to work on.

%%%%%%%%%%%%%%%%

\section*{Acknowledgements}
{We  thank the referee for very constructive comments and suggestions that help to improve our paper significantly.} We are grateful for the discussions with Wei Du and Ran Li. This
research is supported in part by the NSFC of China under grants 11333001,
11173001 and by Strategic Priority Research Program  \textit{The
Emergence of Cosmological Structures}  of the Chinese Academy
of Sciences, grant No. XDB09000000. 
{X.K.L. acknowledges the support from YNU Grant KC1710708 and General Financial Grant from China Postdoctoral Science Foundation with Grant No. 2016M591006.}
Q. W. acknowledges the support from NSFC with Grant No. 11403035.
%The $N$-body simulations are performed on the Shuguang cluster at Shanghai Normal University, Shanghai, China. \textbf{The data analysis is supported by High-performance Computing Platform of Peking University.}
{We acknowledge the 
Shuguang cluster at Shanghai Normal University 
and
the High-performance Computing Platform of Peking University
for providing computational resources.}

\bibliographystyle{aasjournal}
\bibliography{draft_apj_fan}
\end{document}